\documentclass{nature}
\usepackage[utf8]{inputenc} %% For character encoding.
\usepackage[left]{lineno} %% For line numbering.
\usepackage{cite} %% For citation style [1,2,3,4,5] --> [1-5].
\usepackage{graphicx} %% For figures.
\usepackage{url} %% For urls.
\newif\iffinal
\iffinal\renewcommand{\includegraphics}[2][]{}\fi

\title{Core-Collapse Supernova Explosion Theory}
\author{A. Burrows$^{1\star}$ \& D. Vartanyan$^2$}

\begin{document}
\maketitle

%% Affiliation list. List affiliations in order of appearance.
\begin{affiliations}
 \item Department of Astrophysical Sciences, Princeton University, Princeton, NJ 08544, USA
 \item Department of Astronomy, University of California, Berkeley, CA 94720-3411
 \item[$^{\star}$] e-mail: burrows@astro.princeton.edu 
\end{affiliations}

%% Turn on line numbering
%%%%\linenumbers

%% Opening bold paragraph, limited to ~200 words

\begin{abstract}
Most supernova explosions accompany the death of a massive star.  These explosions give birth
to neutron stars and black holes and eject solar masses of heavy elements.  However, 
determining the mechanism of explosion has been a half-century journey of great complexity.
In this paper, we present our perspective of the status of this theoretical quest and
the physics and astrophysics upon which its resolution seems to depend.  The delayed neutrino-heating
mechanism is emerging as a robust solution, but there remain many issues to address, not the least of
which involves the chaos of the dynamics, before victory can unambiguously be declared.
It is impossible to review in detail all aspects of this multi-faceted, more-than-half-century-long
theoretical quest. Rather, we here map out the major ingredients of explosion and the emerging systematics
of the observables with progenitor mass, as we currently see them. Our discussion will of necessity
be speculative in parts, and many of the ideas may not survive future scrutiny. Some statements
may be viewed as informed predictions concerning the numerous observables that rightly exercise
astronomers witnessing and diagnosing the supernova Universe.  Importantly, the same explosion in the inside,
by the same mechanism, can look very different in photons, depending upon the mass and radius of the
star upon explosion.  A 10$^{51}$-erg (one ``Bethe") explosion of a red supergiant with a massive
hydrogen-rich envelope, a diminished hydrogen envelope, no hydrogen envelope,
and, perhaps, no hydrogen envelope or helium shell all look very different, yet might have
the same core and explosion evolution.
\end{abstract}

\section{Core-Collapse Supernova Explosions}
\label{intro}

Stars are born, they live, and they die.  Many terminate their thermonuclear lives after billions 
of years of cooking light elements into heavier elements by ejecting their outer hydrogen-rich
envelopes over perhaps hundreds of years.  In the process, they give birth to compact white dwarf stars, half as massive as the Sun,
but a hundred times smaller.  Such dense remnants cool off over billions of years like dying embers
plucked from a fire.  A subset of these white dwarfs in binary stellar systems will later (perhaps hundreds 
of millions of years) ignite in spectacular thermonuclear explosions, many of these the so-called Type 
Ia supernovae used, due to their brightness from across the Universe, to take its measure. 

However, some stars, those more massive than $\sim$8 M$_{\odot}$,
die violently in supernova explosions that inject freshly synthesized elements, generation 
after generation progressively enriching the interstellar medium with these products of existence.
They too leave behind remnants, but neutron stars and black holes.  The former could become radio 
pulsars, are only the size of a city, and have on average masses 50\% again as massive as the Sun.  
The latter are perhaps a few to ten times more massive than a neutron star, but even more compact and 
more exotic.

The supernova explosions of these massive stars, the so-called core-collapse supernovae (CCSNe),
have been theoretically studied for more than half a century and observationally 
studied even longer. Yet, the mechanism of their explosion has only recently come into sharp 
focus.  A white dwarf is birthed in these stars as well, but before their outer envelope can be ejected
this white dwarf achieves the ``Chandrasekhar mass"\cite{1939isss.book.....C} near $\sim$1.5 M$_{\odot}$.  This mass is gravitationally
unstable to implosion.  After a life of perhaps $\sim$10-40 million years, the dense core of this star 
implodes within less than a second to neutron-star densities, at which point it rebounds like a spherical piston,
generating a shock wave in the outer imploding core.  The temperatures and densities achieved
lead to the copious generation of neutrinos, so the treatment of the neutrinos and their interaction 
with dense matter are of great import.  This ``bounce" shock wave could have been the supernova, but in all
credible models this shock wave stalls into accretion, halting its outward progress.  This is an unsatisfactory 
state of affairs $-$ a supernova needs to be launched most of the time to be consistent with observed rates
and statistics. 

What has emerged recently in the modern era of CCSN theory is that the structure of the progenitor 
star, turbulence and symmetry-breaking in the core after bounce, and the details of the 
neutrino-matter interaction are all key and determinative of the 
outcome of collapse.  Spherical simulations seldom lead to explosion.  Multi-dimensional turbulent convection 
in the core necessitates complicated multi-D radiation(neutrino)/hydrodynamic simulation codes, and these are 
expensive and resource intensive.  It is this complexity and the chaos in the core dynamics after implosion
that has retarded progress on this multi-physics, multi-dimensional astrophysical problem, until now.  
In the first era of CCSN simulations, the state-of-the-art was 
good spherical codes that handled the radiation acceptably.  These models rarely, if ever, exploded. 
Multi-D codes were not yet useful.  Then, two-dimensional (axisymmetric) codes arrived, captured 
some aspects of the overturning convection about which 1D models are mute, but were 
slow $-$ only a few runs could be accomplished per year. This era was followed by the advent of some 
3D capability, and at the same time many 2D runs could be performed to map out some of parameter 
space and gain intuition concerning the essential physics and behavior.  We are now in the era 
of multiple 3D simulations per year, wherein we can explore core dynamics and explosion in the full
3D of Nature without the fear that a mistake in a single expensive run which could take a year on a supercomputer 
would set us back.  This progress has been enabled by the parallel expansion of computer power over the decades.
It is the pivotal role of multi-dimensional turbulence and the breaking of spherical symmetry in the mechanism 
of explosion itself, coupled with the driving role of neutrino heating, that necessitated the decades-long numerical 
and scientific quest for the mechanism of core-collapse supernovae.  What Nature does effortlessly 
in a trice has taken humans rather longer to unravel. 

However, there are now strong recent indications that the dominant explosion mechanism and rough 
systematics of the outcomes with progenitor star are indeed yielding to ongoing multi-pronged international 
theoretical efforts.  Moreover, code comparisons are starting to show general concordance\cite{oconnor_code2018}. 
Many recent multi-dimensional simulations employing sophisticated physics 
and algorithms are exploding naturally and without artifice.  These include those from our group  
\cite{radice:17,burrows:18,vartanyan2018a,vartanyan2019,burrows_low_2019,2019MNRAS.490.4622N,2020MNRAS.491.2715B,2020MNRAS.492.5764N}, 
using the state-of-the-art code F{\sc{ornax}}\cite{fornaxcode:19}, and those from others 
\cite{lentz:15,melson:15b,2015ApJ...801L..24M,jms2016,TaKoSu16,muller2017,oconnor_couch2018,kuroda:18,2019ApJ...873...45G,2020ApJ...896..102K}.  
Neutrino heating in the so-called gain region behind a stalled shock, aided by
the effects of neutrino-driven turbulence and spherical symmetry breaking, together seem, in broad outline,
to be the agents of explosion for the major channel of CCSNe.  Other subdominant channels might be thermonuclear 
(what do the terminal cores of $\sim$8-9 M$_{\odot}$ stars actually do?\cite{jones_2016,2019PhRvL.123z2701K,2019A&A...622A..74J}, but see\cite{2019ApJ...886...22Z}) 
or magnetically-driven (so-called ``hypernovae", $\sim$1\%\cite{burrows2007_mag,mosta2015};   
long-soft gamma-ray bursts, $<0.1$\%).  And indeed, there remain for the neutrino mechanism numerous interesting 
complications concerning nuclear and neutrino physics, the progenitor structures, and numerical challenges 
to be resolved before this central problem can be retired.

\section{How do Core-collapse Supernovae Explode?}
\label{mech}

It is generally agreed that the stall of the roughly spherical bounce shock wave sets up a quasi-hydrostatic 
structure interior to it that accretes the matter falling through the shock from the outer core that 
is still imploding \cite{Bethe:1990mw,2005NatPh...1..147W,janka2012,burrows2013}.  The rate of
accretion ($\dot{M}$) through the stalled shock and onto the inner core is an important evolving quantity that depends
essentially upon the density structure of the progenitor's core just prior to Chandrasekhar instability 
(see Figure \ref{fig:rho}) and determines much of what follows.  The core is so dense and the neutrino particle energies are so high
(10's to 100's of MeV, million electron volts) that this structure interior to $\sim$10$^{11}$ g cm$^{-3}$ is opaque to neutrinos 
of all species - the structure is a ``neutrino star" with  ``neutrinosphere" radii that depend upon the 
neutrino species and particle energy, but are initially $\sim$30-60 kilometers (km) in radius.  The bounce
shock initially forms in the deeper neutrino-opaque region, but as it emerges quickly (within milliseconds) 
to larger radii and lower densities a burst of electron neutrinos ($\nu_e$) is generated.  It is this burst
that saps energy from the shock and leads to its stalling into accretion.  A secondary cause of its stalling is 
the shock dissociation of the infalling nuclei into nucleons. This effect lowers the effective ``$\gamma$" 
of the gas that connects internal thermal energy with pressure by diverting energy into nuclear 
breakup, thereby channeling less efficiently the gravitational energy otherwise available to provide pressure 
support for the shock. The stalled shock radius initially hovers around 100-200 km. Just interior to the shock is 
the semi-(neutrino)transparent ``gain region" where the ``optical" depth to neutrinos is $\sim$0.1. This region 
surrounds the neutrinospheres through which most of the ongoing prodigious neutrino emissions emerge
and these bound the inner dense core containing most of the PNS (``proto-neutron star") mass.  This quasi-stable proto-neutron star 
bounded by the stalled shock fattens by accretion and shrinks by neutrino loss.  The neutrino emissions
are powered by thermal diffusion from the interior and the gravitational power of accretion. The goal of theory 
is to determine how the shock is reenergized and launched into explosion, leaving behind the bound 
neutron star.  The explosion is of the mantle of the PNS approximately exterior to the neutrinospheres, 
and the bound inner material must be left behind. 
 
If there were no ongoing accretion, then neutrino heating in the gain region behind the shock 
wave would be more than sufficient to power an exploding shock.  There would be no tamping accretion
ram pressure and neutrino heating by $\nu_e$ and $\bar{\nu}_e$ neutrino absorption on 
free neutrons and protons in the gain region would easily power a dynamical outflow. 
This is similar to a thermal wind.  However, the accretion ram and neutrino heating are competing to determine 
instability to explosion, with the added complication that accretion is also powering a changing fraction of 
the driving neutrino luminosities. The explosion is akin to a bifurcation between quasi-stationary accretion 
and explosion solutions, with control parameters related to the accretion rate and the neutrino luminosities 
(for two), but a simple analytic explosion condition in the context of realistic simulations has proven elusive.      
Hence, detailed simulations are required.

What has emerged is that only those progenitor models with very steep outer density 
profiles that translate into rapidly decreasing post-bounce accretion rates can explode in 
spherical symmetry (1D) via the neutrino mechanism.  Among 
the representative progenitor models shown in Figure \ref{fig:rho}, only the 9 M$_{\odot}$ star comes close to fitting that 
description.  However, not even it explodes in our 1D simulations. Multi-D seems required. {Classically, the 8.8 M$_{\odot}$ 
model of Nomoto\cite{nomoto:84} explodes spherically, as do a few others with similar very steep outer 
density profiles\cite{kitaura,radice:17}.  However, due to new ideas concerning the character of thermonuclear burning 
and electron capture in such compact cores, this lowest-mass progenitor region is undergoing a modern reappraisal
\cite{jones_2016,2019PhRvL.123z2701K,2019A&A...622A..74J}. Note that there are strong arguments in tension with
this alternate perspective\cite{2019ApJ...886...22Z}.} 
The current explosion paradigm for most massive stars is gravitational-energy sourced, neutrino-driven, and turbulence-aided,
and we now summarize some of what we have learned concerning the roles of various specific physical effects.

%Using both 1) long-term 
%2D F{\sc{ornax}}\cite{fornaxcode:19} code simulations for the representative broad suite of non-rotating 
%progenitor models\cite{sukhbold2018} from 9 to 27 M$_{\odot}$ shown in Figure \ref{fig:rho} and 
%2) multiple 1D, 2D, and 3D simulations we have published over the last several years
%\cite{radice:17,burrows:18,vartanyan2018a,vartanyan2019,burrows_low_2019,2020MNRAS.491.2715B},

\subsection{Efficiency}
\label{eff}

Since a hot and lepton-rich PNS radiates $\sim$3$\times$10$^{53}$ ergs in neutrinos as it transitions into a tightly
bound, cold neutron star and supernova explosion energies are ``typically" one Bethe, 
it is often stated that the neutrino mechanism of core-collapse explosion is one of less than 1\% tolerances.
This is not true.  During the 100s of milliseconds to few-second timescales after bounce over which 
the neutrino heating mechanism operates, the efficiency of energy deposition in the gain region, the fraction 
of the emitted energy absorbed there, is $\sim$4$-$10\%, far higher. Most of the binding energy 
of the neutron star is radiated over a period of a minute\cite{bhf1995}, after the phase during which 
we think the explosion energy is fully determined.

\subsection{Turbulent Convection}
\label{turb}

\begin{figure*}[ht!]
    \begin{center}

    \includegraphics[width=0.95\linewidth]{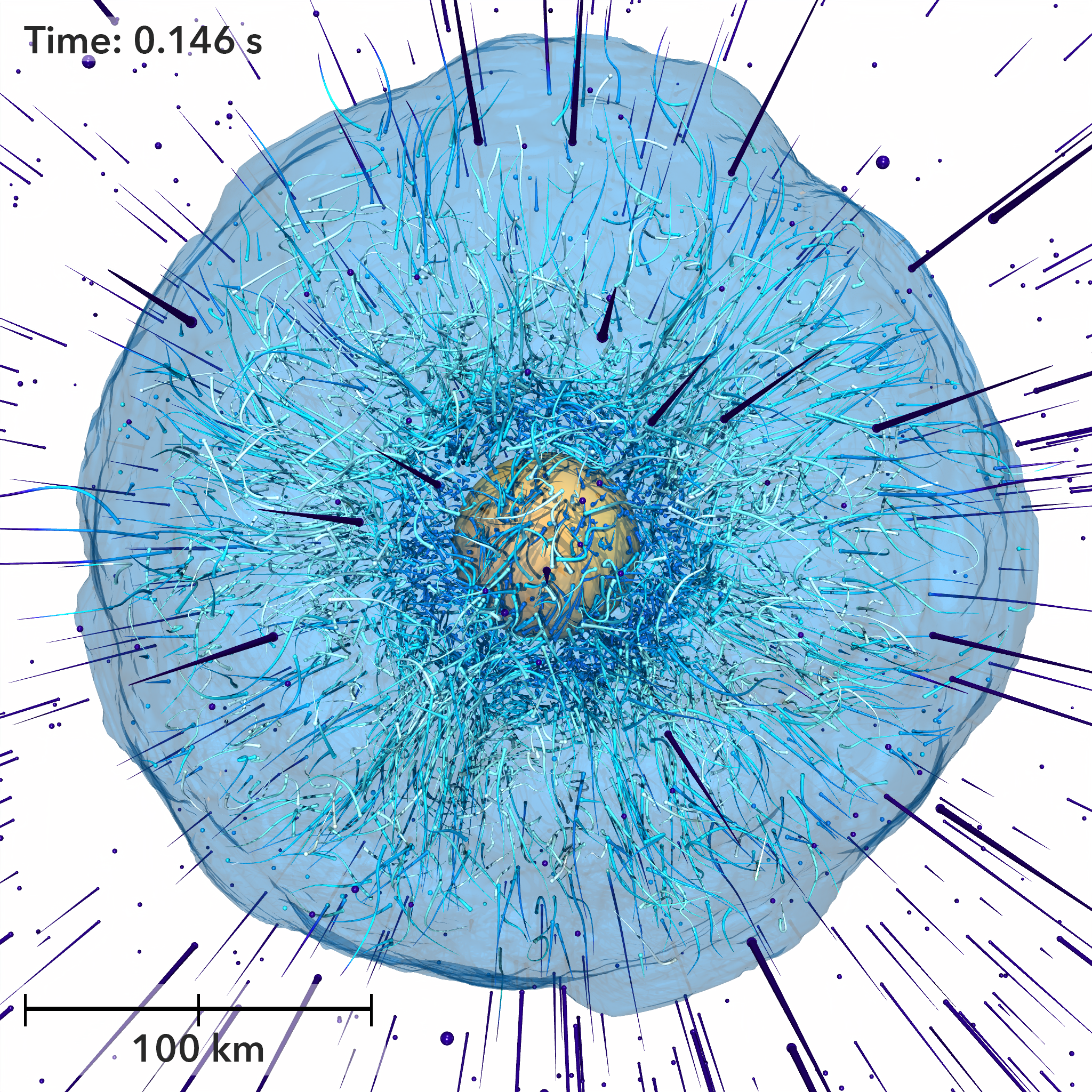}

    \end{center}
    \caption{{\bf Inner matter trajectories as the explosion is about to launch.}  Shown are
the interiors of an explosion only $\sim$150 milliseconds
after core bounce (vertical physical scale $\sim$350 kilometers). At this time
the shock wave is at $\sim$150 kilometers, just before explosion.
The inner ball is the newly-birthed proto-neutron star (PNS) (rendered as an isodensity
surface at 10$^{11}$ g cm$^{-3}$, colored by Y$_e$), surrounded by swirling, turbulent matter, most
of which will settle onto the PNS.  The trajectories depict the recent 5 milliseconds in the positions of
individual accreted matter elements.  They are colored by local entropy. The turbulence of this inner region
is manifest.
\label{fig:tracers}}
\end{figure*}

Turbulence is fundamentally a multi-dimensional phenomenon and can't be manifest in spherical 
(1D) symmetry (and, therefore, in 1D simulations).  The turbulence in the gain region 
interior to the stalled shock is driven predominantly by the neutrino heating 
itself, which produces a negative entropy gradient unstable to overturn. 
This is similar to boiling water on a stove, via absorptive heating from below\cite{bhf1995}. 
Figure \ref{fig:tracers} depicts the inner turbulent convective region early after 
bounce before explosion, showing accreted matter tracers swirling randomly about the PNS core. A larger 
neutrino heating rate will increase both the vigor of the turbulence and the entropy of this 
mantle material. The matter that accretes through the shock on its way inward to the PNS during the 
pre-explosion phase contains perturbations\cite{2006ApJ...652.1436F,2014ApJ...795...92C,2015ApJ...799....5C,muller_janka_pert,MuMeHe17} 
that arise during pre-collapse stellar evolution which will seed the convective 
instability. The larger and more prevalent these seeds the quicker the 
turbulence grows to saturation and in vigor. One feature of turbulence is turbulent pressure.
The addition of this stress to the gas pressure helps push the shock to a larger stalled 
shock radius. This places matter in more shallow reaches of the gravitational potential well 
out of which it must climb and helps to overcome the subsequently smaller ram pressure 
due to infalling matter from the outer core still raining in.  The turbulence
also forces the accreted matter to execute non-radial trajectories as it settles, increasing the time
during which it can absorb neutrino energy before settling on the PNS and, hence, the average entropy 
that can be achieved in the gain region\cite{murphy:08}.  Therefore, through the combined 
agency of both neutrino heating and neutrino-driven turbulence, the quasi-stationary structure 
that is the PNS, plus mantle, plus stalled shock wave is more likely to reach a critical 
condition wherein the steady infalling solution bifurcates into an explosive one. The huge binding energy
accumulated in the PNS does not need to be overcome $-$ only its mantle (and with it the rest of the star)
needs to be ejected.  

Moreover, the turbulent hydrodynamic stress is anisotropic, with its largest component 
along the radial direction\cite{murphy:08}. {Turbulent magnetic stress might also be a 
factor\cite{2020MNRAS.498L.109M}.}  Importantly, turbulence is more effective at using energy to
generate stress/pressure than a gas of nucleons, electrons, and photons. 
As much as $\sim$30$-$40\% of the stress behind the stalled shock when the turbulence 
is fully developed can be in turbulent stress. Hence, partially channelling gravitational energy of 
infall into turbulence instead of into thermal energy helps support and drive the shock more efficiently
\cite{2019MNRAS.490.4622N}.

\subsection{Neutrino-Matter Interactions}
\label{netrino}

The predominant processes by which energy is transferred from the radiated neutrinos to the matter behind the 
shock in the gain region are electron neutrino absorption on neutrons via $\nu_e + n \rightarrow e^- + p$, anti-electron 
neutrino absorption on protons via $\bar{\nu}_e + p \rightarrow e^+ + n$, and inelastic scattering of 
neutrinos of all species off of both electrons and nucleons.  The two super-allowed charged-current absorption
reactions dominate and provide a power approximately equal to the product of the neutrino luminosity and the neutrino optical 
depth in the gain region.  The latter can be $\sim$10\%.  Therefore, the higher the luminosity and/or 
absorption optical depth the greater the neutrino power deposition, which can reach levels of many Bethes per second.
Upon explosion most of this power goes into work against gravity and only a fraction of the deposited energy 
is left as the asymptotic blast kinetic energy.  This is very qualitatively similar to a 
thermally-driven wind, for which the energy at infinity scales with the binding energy of the ejecta. Therefore, 
the stellar binding energy of the ejected mantle might approximately set the scale of the supernova explosion energy.
We discuss aspects of this paradigm in \S\ref{energy}.

At higher mass densities ($\rho$) above 10$^{11}$ $-$ 10$^{12}$ g cm$^{-3}$, nucleon-nucleon interactions 
introduce correlations in density and spin. Such non-Poissonian correlations modify the scattering 
and absorption neutrino-matter interaction rates (generally suppressing them), hence, they affect the emergent neutrino 
luminosities\cite{1998PhRvC..58..554B,sawyer1999,reddy1999,burrows:06,roberts2012,2020PhRvC.101b5804F}. This is relevant to the instantaneous power
deposition in the gain region, and, hence, the neutrino-driving mechanism itself. These many-body effects 
increase with density and some of the associated correction factors have been estimated\cite{roberts_reddy2017,PhysRevC.95.025801}
to be of order 10-20\% at 10$^{12}$ g cm$^{-3}$, near and just interior to the neutrinospheres. However, such 
corrections depend upon a detailed and self-consistent treatment of the opacities along with the nuclear 
equation of state (EOS), and this goal has yet to be achieved. Nevertheless, using scattering suppression factors 
researchers\cite{horowitz:17,burrows:18} have shown that these effects can facilitate
explosion.  They do this by decreasing the opacities, thereby increasing the neutrino loss rates.  This leads to a more
rapid shrinking of the PNS, which due to consequent compression heats the neutrinosphere regions.  This increases
the mean energy of the emitted neutrinos.  Since the neutrino absorption rates via the charged-current
reactions quoted aobve increase approximately as the square of the neutrino energy
and the luminosities themselves are elevated, the neutrino power deposition in the gain region is augmented, thereby
facilitating explosion.  The effect is not large, but when an explosion is marginal it can be determinative. 

During the early collapse phase, increasing densities lead to increasing electron Fermi energies 
and higher electron capture rates on both free protons and nuclei.  Electron capture decreases
the electron fraction (Y$_e$, the ratio of the electron density to the proton plus neutron density) 
of the infalling gas, and this decreases the electron pressure. 
A decrease in the electron pressure slightly accelerates the infall and the mass accretion rate 
($\dot{M}$) versus time.  As already stated, $\dot{M}$ after inner core bounce is a key parameter determining,
among other things, the accretion ram pressure external to the shock and the accretion component of the neutrino luminosities.
Therefore, the rate of capture on infall can affect explosion timing and, perhaps, its viability.
The effect is not large, but when things are marginal altering the evolution of $\dot{M}$ can be
important.  However, the capture rate on the mix of nuclei in the imploding core is not 
known to better than perhaps a factor of five \cite{2003PhRvL..90x1102L,2010NuPhA.848..454J,2012ApJ...760...94L}.
Hence, clarifying this important issue remains of interest to modelers.

Finally, the energy transfer to matter via inelastic scattering off electrons and nucleons provides
a subdominant component of the driving heating power behind the shock wave.  The effect may be
only 10\%-15\%, but, again, when the core teeters on the edge of explosion such effects matter.
Neutrino scattering off electrons (akin to Compton scattering, but for neutrinos) results in a large 
energy transfer, but has a small rate.  Energy transfer to the heavier nucleons is small, but the scattering 
rate is large.  The net effect results in comparable matter heating rates for both effects, with a slight advantage 
to neutrino-nucleon scattering\cite{burrows:18}. However, calculating such spectral energy redistribution
is numerically difficult, and represents one of the major computational challenges in the 
field\cite{1993ApJ...410..740M,burrows:06,chimera}.

%%%% EOS??

\subsection{Explosion}
\label{explosion}

\begin{figure*}[ht!]
    \begin{center}
    \includegraphics[width=0.9\linewidth]{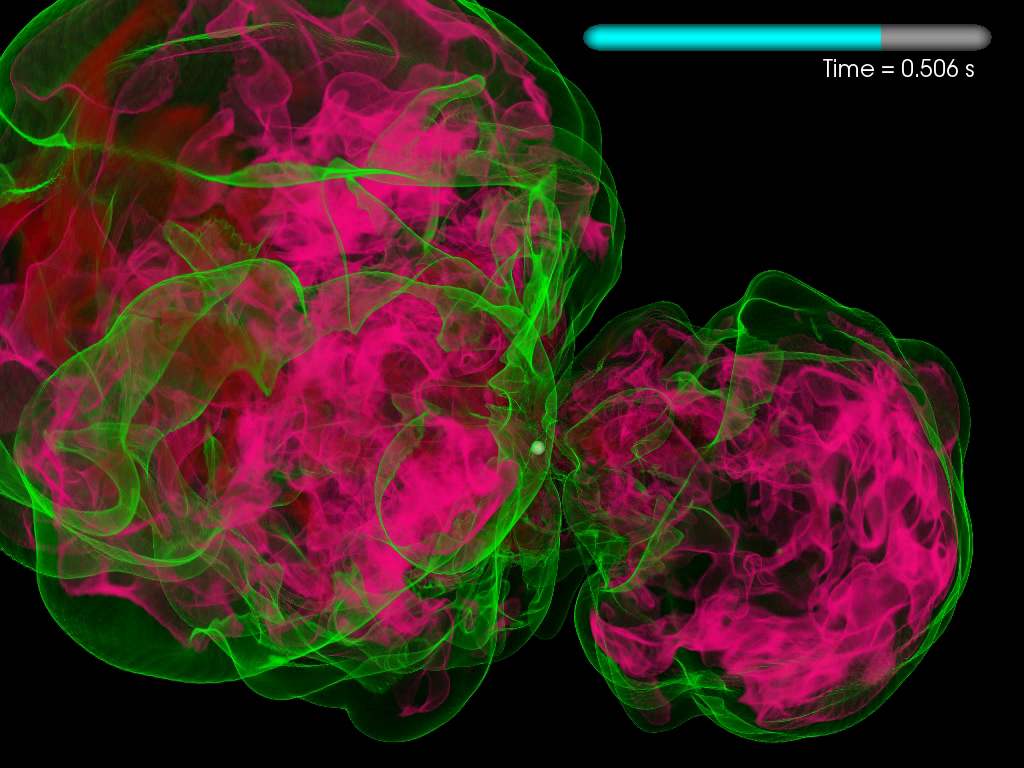}
    \end{center}
    \caption{{\bf Early 3D explosion of the core of a 16-M$_{\odot}$ star using F{\sc{ornax}}.} Portrayed is a still near 500 milliseconds 
after core bounce at nuclear densities.  The red is a volume rendering of the high-entropy of the ejecta in 
the neutrino-heated bubbles that constitute the bulk of the volume of the exploding material.  The green surface is 
an isoentropy surface near the leading edge of the blast, the supernova shock wave.  Note the asymmetric, though roughly
dipolar, character of the explosion and the pinched ``wasp-waist" structure of the flow between the lobes. The dot
at the center in the newly born neutron star. In this model, as in many others, there is clearly simultaneous 
accretion at the waist, while there is ejection in the wide-angle lobes. Simultaneous accretion in one sector 
during concommitant explosion elsewhere maintains the driving neutrino luminosity
and is a signature of the useful breaking of spherical symmetry possible in multi-dimensional flow.  
This contrasts sharply with the artificially enforced situation in 1D/spherical simulations.
Simulation performed by the Princeton supernova group\cite{vartanyan2019}. \label{fig:16.550}}
\end{figure*}

The stalled shock radius can be decomposed into spherical harmonics in solid angle. The onset of 
the explosion of a stalled accretion shock is a monopolar instability in the quasi-spherical shock. 
However, approximately when the monopole becomes unstable, the dipole often seems to as well\cite{2013ApJ...765..110D,vartanyan2019}.  
Therefore, the explosion picks an axis, seemingly at random for a non-rotating progenitor, and the blast has a dipolar 
structure with a degree of asymmetry that seems to be low for quickly exploding models and larger for those 
whose explosion is more delayed.  Figure \ref{fig:16.550} depicts an example blast structure manifesting 
such a dipole\cite{vartanyan2019}. Generally, but not completely reliably, the lower-mass progenitors
(such as a 9-M$_{\odot}$ progenitor\cite{burrows_low_2019,2020MNRAS.491.2715B}) explode earlier and before 
turbulence is vigorous and the more massive progenitors seem to explode
later and after turbulence has achieved some vigor.  Hence, the latter generally, though not every time, 
explode more asymmetrically, with a larger dipolar component\cite{2019MNRAS.489.2227V}. The chaos of the turbulence 
makes the outcome stochastic, so that the direction of explosion is not easily predicted. Importantly, the
chaos of the turbulent flow will result in distribution functions of explosion times, directions, explosion energies, 
explosion morphologies, residual neutron star masses, $^{56}$Ni yields, general nucleosynthesis, and kick velocities,
etc., even for the same star. It is not even known whether those functions are broad or narrow for a given star.

In addition, exploding more along an axis, as depicted in Figure \ref{fig:16.550}, allows the flow external to the 
shock to wrap around the prevailing axis and accrete along a pinched waist in an equatorial structure.  This breaking of 
symmetry, impossible in spherical symmetry, allows simultaneous accretion and explosion. Whereas a 1D explosion
by its nature turns accretion off in all directions, and thereby throttles back the accretion component of
the driving neutrino luminosity, in multi-D the accretion component of the luminosity can be maintained.
Hence, the breaking of spherical symmetry supports the driving luminosity and facilitates explosion, just as 
it is getting started.  This symmetry breaking is an important aspect of viable explosion models and is impossible 
in spherical models.  Nature unchained to manifest overturning instability leading to turbulence employs
this freedom to facilitate explosions that might be thwarted in 1D.  Both the turbulent stress and the option 
of simultaneous accretion in one direction while exploding in another are important features of the CCSN explosion 
mechanism.

Convection in the progenitor star upon collapse will create perturbations in velocity, 
density, and entropy that seed overturn and turbulence in the post-shock matter exterior to the inner 
PNS and the neutrinospheres. The magnitude of these perturbations generally increases with progenitor mass,
but their true character is only now being explored in detail.  Recently, a number of groups
have embarked upon 3D stellar evolution studies during the terminal stages of massive stars
\cite{CoChAr15,jones_2016,2016ApJ...822...61C,muller2017,2019MNRAS.484.3307M,2019A&A...622A..74J}. 
The potential role of aspherical perturbations in the progenitor models in inaugurating
and maintaining turbulent convection behind the stalled shock wave
is an active area of research\cite{2015ApJ...799....5C,muller2017,burrows:18,vartanyan2018a,2019MNRAS.484.3307M}
and these studies might soon reveal the true nature of accreted asphericities and their spatial distribution.
It might also be that low-order modes in the progenitors would naturally result in angular asymmetries 
in the mass accretion through the shock and provide a path of least resistance that would (however randomly) 
set the explosion dipole and direction, whatever its magnitude.  Such low accretion-rate paths might actually facilitate
explosion in circumstances when it would otherwise be problematic.

Another convective phenomenon that can help achieve the critical condition for explosion is proto-neutron-star convection.
This is not the neutrino-heating-driven convection in and near the gain region just behind the stalled shock,
but overturn driven by electron lepton loss from beneath the neutrinospheres.  As electron neutrinos are liberated from the inner
PNS mantle around a radius of $\sim$20 km, the resulting negative Y$_e$ gradient is convectively unstable.
This is akin to instabilities in stars due to composition gradients.  All proto-neutron stars show this instability,
which lasts for the entire duration of PNS evolution and likely continues long after (many seconds to one minute)
the explosion is launched (if it is). PNS convection\cite{1996ApJ...473L.111K,dessart_06,2020MNRAS.492.5764N} accelerates energy loss (particularly via $\nu_{\mu}$,
$\nu_{\tau}$, $\bar{\nu}_{\mu}$, and $\bar{\nu}_{\tau}$ neutrinos) and electron lepton loss in the PNS, thereby accelerating 
core shrinkage.  In a manner similar to the many-body effect, such core shrinkage leads to higher neutrinosphere 
temperatures and a stronger absorptive coupling to the outer gain region.

\begin{figure*}[ht!]
    \begin{center}
    \includegraphics[width=0.9\linewidth]{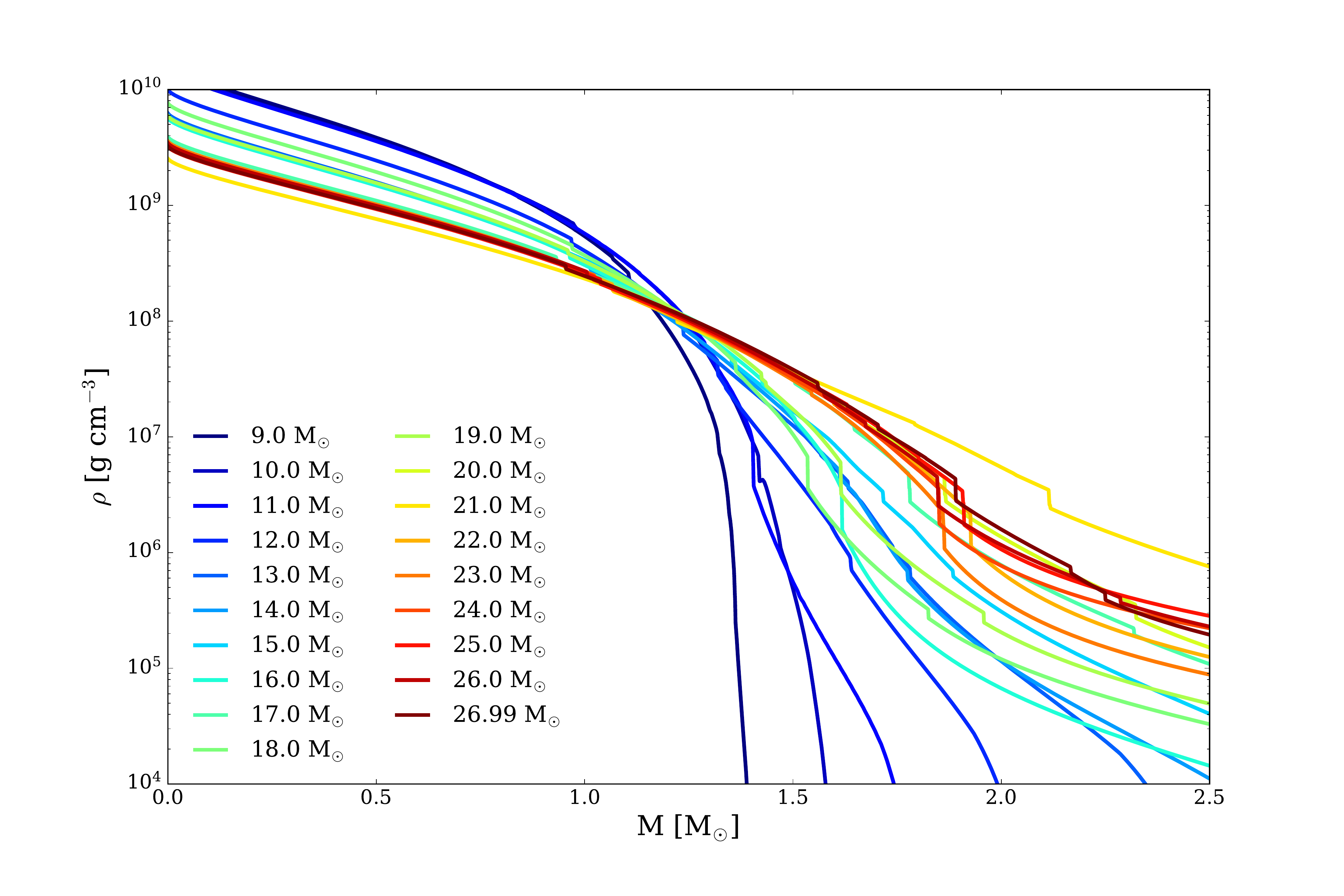}
    \end{center}
    \caption{{\bf Progenitor mass density profiles.} Plotted are the mass density (in g cm$^{-3}$) versus interior mass 
(in M$_{\odot}$) profiles of the cores of the progenitor massive stars used as initial conditions for the 
supernova simulations we highlight in this paper. The associated spherical stellar evolution models were 
calculated by Sukhbold et al. \cite{sukhbold:16,sukhbold2018} up to the point of core collapse, at which point they were mapped 
into our supernova code F{\sc{ornax}} \cite{fornaxcode:19}. \label{fig:rho}}
\end{figure*}

We end this section by emphasizing that the most important determinant of explosion, all else being equal, 
is the mass density structure of the unstable Chandrasekhar core. The density profile translates 
directly into the mass accretion rate after bounce and this determines both the accretion tamp
and the accretion component of the driving neutrino luminosity.  Figure \ref{fig:rho} provides an 
example set\cite{sukhbold2018} of density profiles ($\rho(r)$) from 9 M$_{\odot}$ to 27 M$_{\odot}$.  This set
spans most (but not all) massive stars that give birth to CCSNe.  There are a few trends in $\rho(r)$ worth 
noting.  First, the lowest-mass massive stars generally have slightly higher central densities and steeper 
outer profiles and the higher-mass massive stars have lower central densities and significantly shallower 
outer density profiles. However, the trend in the slope of the outer density profiles is not strictly monotonic
with progenitor mass, with some ``chaos" in the structures.  Ambiguities in the handling of convection, 
overshoot, doubly-diffusive instabilities, and nuclear rates have led to variations from modeler to modeler
in progenitor stellar models up to collapse that have yet to converge. Furthermore, the effects of fully 3D 
stellar evolution and rotation have not yet been fully assessed.  Therefore, the summary behavior depicted 
in Figure \ref{fig:rho} is provisional.  

Given these caveats, important general insights are emerging.  The first is that the silicon/oxygen shell
interface in progenitors (seen for many models in Figure \ref{fig:rho}) constitutes an important 
density jump, which if large enough can kickstart a model into explosion. In many of our 
models, the shock is ``revived" upon encountering this interface\cite{2011ApJ...730...70O,2020MNRAS.491.2715B}.  
The associated abrupt drop in accretion rate and inhibiting ram pressure at the shock upon the accretion of 
this interface is not immediately followed by a corresponding drop in the driving 
accretion luminosity. This is due to the time delay between accretion to the shock at 100$-$200 km 
and accretion to the inner core where the gravitational energy is converted into useful 
accretion luminosity. This time delay pushes the structure closer to the critical point 
for explosion.  However, sometimes the density jump is not sufficiently large and 
its magnitude in theoretical stellar models has not definitively been pinned down.

Second, the very steep density profiles seen for 
the lower-mass massive stars lead to earlier explosions. However, the associated steeply decreasing rates of
accretion also result in less mass in the gain region and a lower optical depth to the emerging neutrino fluxes.
The lower absorption depths in the mantle times the lower neutrino luminosities lead to lower driving powers
in the exploding mantle. This results in lower explosion energies generically under the neutrino heating paradigm
for those stars with steep outer density profiles.  Conversely, those stars with shallow density profiles, more
often the more massive CCSN progenitors, generally explode later.  However, their shallow mass profiles result
in more mass in the gain region with a greater optical depth.  The larger depth times the larger accretion 
luminosities lead to greater driving neutrino power deposition.  The net effect is often higher asymptotic
supernova explosion energies.  Hence, with exceptions, state-of-the-art models suggest that the explosion energy
is an increasing function of progenitor mass and the shallowness of the outer density profile 
of the initial core.  In addition, ``explodability" does not seem to be a 
function of ``compactness"\cite{2011ApJ...730...70O,burrows:18,2020ApJ...888...91E} 
(a measure of the ratio of progenitor interior mass to radius), with both high and 
low compactness models exploding.  It had been suggested that only low-compactness structures
exploded.  Not only does this not seem to be true, but it seems that only the higher compactness models can result in 
explosion energies near the canonical one Bethe. It may be, however, that very high compactness structures
have outer mantle binding energies for which the neutrino mechanism can't provide sufficient driving power.  These 
objects may lead either to weak explosions or fizzles, with many of these leading to black holes (and not neutron stars).
In fact, the gravitational binding energy of the mantle of the Chandrasekhar core may set the scale of the explosion energy, 
and if too high might thwart explosion altogether.  This topic deserves much more attention.

%As should now be clear, a multitude of effects can be of importance in determining the viability, character, and 
%strength of a CCSN explosion.  It is this complexity that must be factored into simulations before an assessment 
%of the dynamical true outcome can be ascertained. This is one major reason that progress on this multi-physics, 
%multi-dimensional, multi-decade puzzle has, until recently and however inexorably, been slow.

%If this can be demonstrated, then the overall problem of the mechanism of supernova explosions
%might at the zeroth or first-order level be considered solved, the rest being details.
%These details include the mapping of progenitor to outcomes (energy, NS mass, nucleosynthesis,
%morphology, BHs(?), pulsar kicks, pulsar and magnetar B-fields, spin rates, nuclear EOS.

\section{Supernova Energies}
\label{energy}

Two-dimensional (axisymmetric) and three-dimensional simulations do not behave the same.  The axial constraint
and artificial turbulent cascade of the former compromise the interpretation of the results.  However, 2D simulations
do allow the breaking of important symmetries and overturning motions and are less expensive to perform.  Importantly, 
due to their much lower cost, 2D numerical runs can easily be carried out to many seconds after bounce, something that 
we and others have found is required for many stars to asymptote to final blast kinetic energies in the 
context of the neutrino mechanism\cite{bmuller_2015}.  So, in order to get a bird's-eye view of some of the systematic
behavior with progenitor mass, we have conducted for this paper a suite of longer-term 2D models using the 
stellar models of Sukhbold et al.\cite{sukhbold2018} as starting points.  For this collection, we have found
usually that when a 2D model explodes, its more realistic 3D counterpart does as well, and when it doesn't neither 
does the 3D simulation.  In our experience, this is usually, but not always, the case, though there is
some disagreement on this in the literature\cite{2012ApJ...755..138H,2013ApJ...765..110D,hanke:13,2013ApJ...775...35C,summa:16,jms2016,oconnor_couch2018,summa:18}.In our recent set of models, it is only the 12-M$_{\odot}$ and 15-M$_{\odot}$ models that do not explode. The 2D models generally seem 
to explode a bit earlier than the 3D models. For instance, the 20-M$_{\odot}$ and 25-M$_{\odot}$ stars explode $\sim$100 ms
and $\sim$50 ms later in 3D.  Also, on average the more massive progenitor models explode later.  Nevertheless, the shock is 
(re)launched, if it is, between $\sim$150 and $\sim$400 ms after bounce for all these 2D exploding models.  This timescale
depends, no doubt, upon simulation details (microphysics, resolution, algorithms, etc.), as well as the character of
the seed perturbations.  For these simulations, we did not impose extra perturbations and left the inauguration
of the initial overturning instabilities to numerical noise.

\begin{figure*}[ht!]
    \begin{center}
    \includegraphics[width=0.75\linewidth]{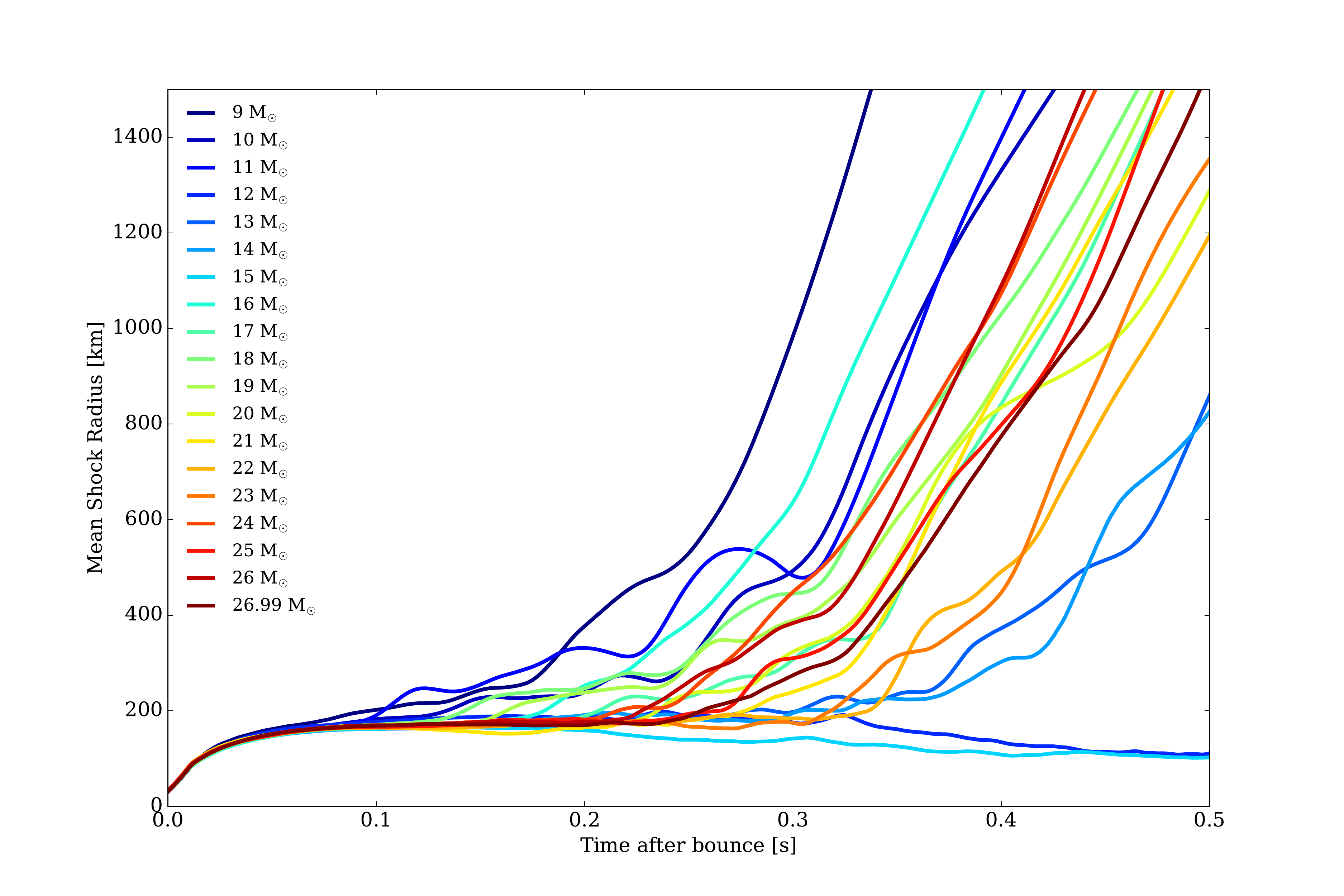}
    \includegraphics[width=0.75\linewidth]{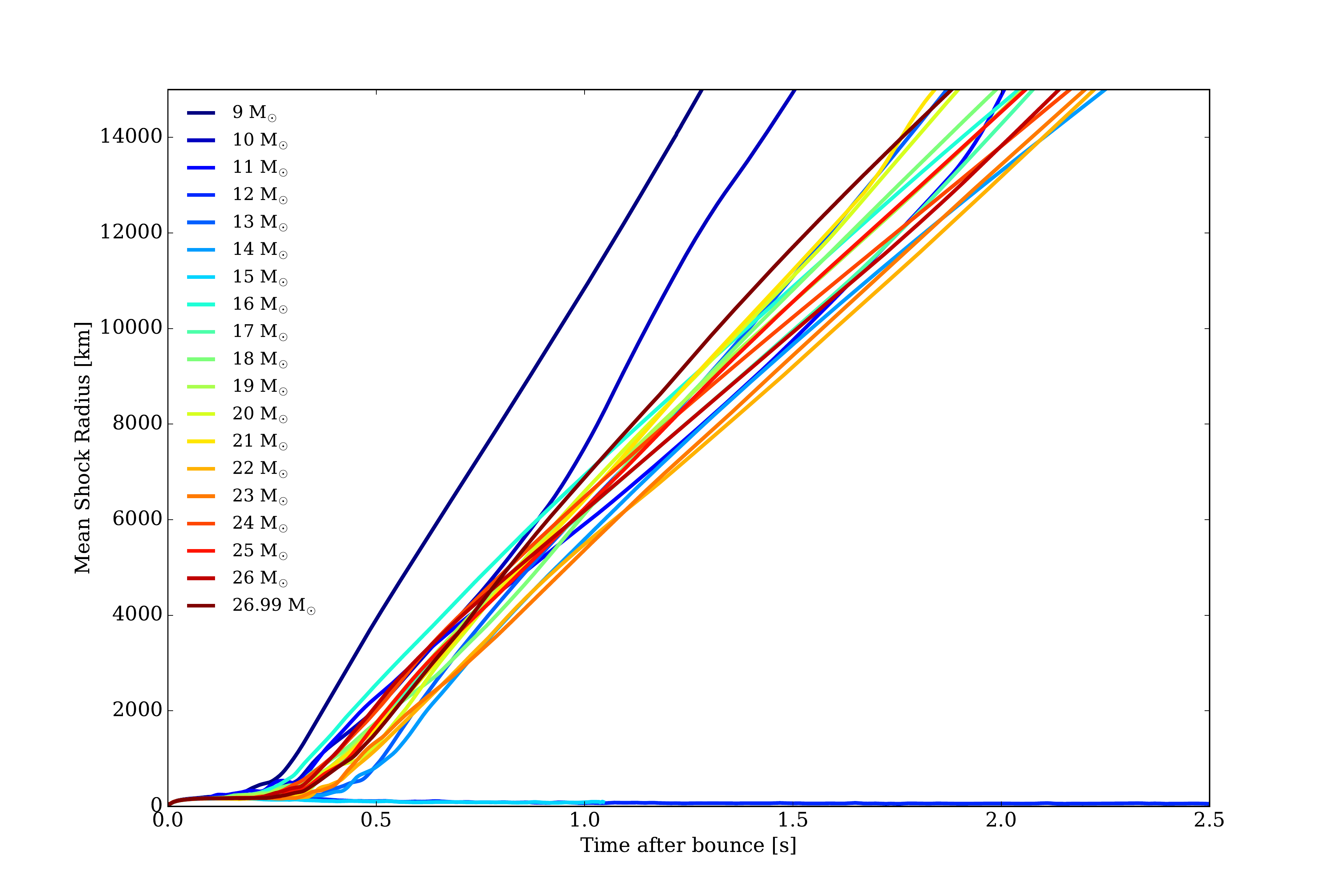}
    \end{center}
    \caption{{\bf Mean shock radii of 2D models.} Depicted are the angle-averaged shock radii (in kilometers) of the 
2D model suite calculated for this paper versus time (in seconds) after bounce.  Most of the models explode, while the 12- and 15-M$_{\odot}$ progenitor
structures do not.  The top panel shows the behavior during the first half second after bounce and in the inner 1500 kilometers, with models 
exploding (when they do) between $t = 0.15$ and $t = 0.4$ seconds.  The bottom panel portrays the shock motion
on a larger physical scale (15000 kilometers) and to latter times. Many of the models were, in fact, carried
to $\sim$4.5 seconds after bounce. The mean shock speeds become rather stable, with values between either 10000 or 15000
km s$^{-1}$ for most of the simulation. The models were conducted on grids from 30,000 to 100,000 kilometers,
with the smaller values for the smaller-mass progenitors.  See the caption to Table \ref{table1} for specifics. \label{fig:rshock}}
\end{figure*}

\begin{figure*}[ht!]
    \begin{center}
    \includegraphics[width=0.9\linewidth]{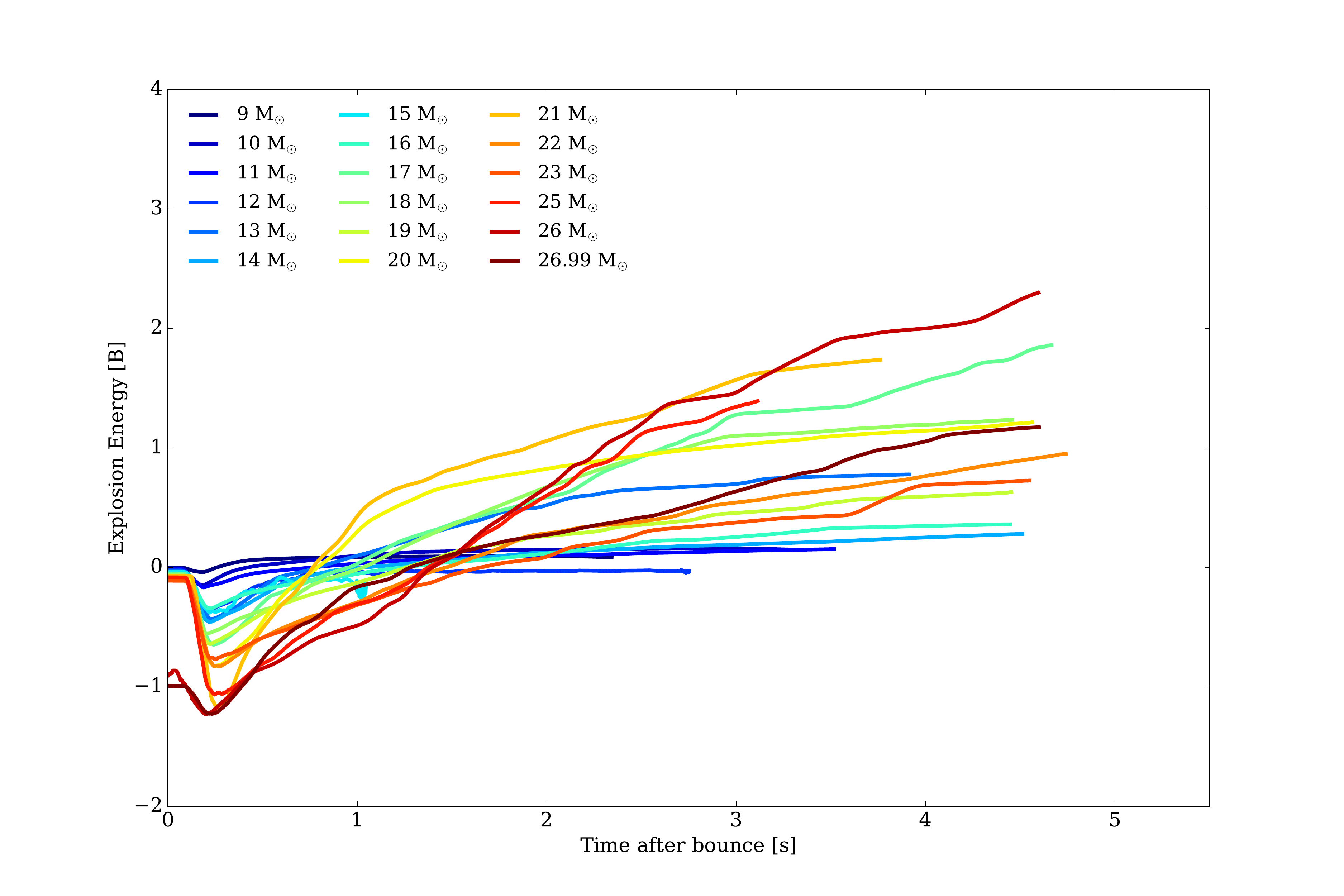}
    \end{center}
    \caption{{\bf The evolution of the total explosion energy (in Bethes) with time 
(in seconds after bounce).}  As the figure indicates, many models start bound (negative energies), even though their shocks have been launched.  
It can take more than one second for some to achieve positive energies, the true signature of an explosion.  Moreover,
as shown on this figure, it can take $\sim$4$-$5 seconds for the supernova energy to asymptote, and some take longer than that.
All the more massive exploding models take this longer time, and they generally achieve the highest supernova energies.
The lower-mass massive progenitors asymptote earliest at generally, though not universally, lower supernova energies.
In addition, though a model might explode late, it can still achieve a higher explosion energy than those that
explode early.  Hence, the time of explosion is not indicative of its eventual vigor. Note that the 12-M$_{\odot}$ 
and 15-M$_{\odot}$ stars in this investigation do not explode.
\label{fig:energy}}
\end{figure*}

Figure \ref{fig:rshock} portrays the development of the mean shock radius for all the models of this study. The left-hand panel
shows the launch phase, while the right-hand panel provides a later, larger-scale glimpse.  The mean shock speeds settle between
10,000 km s$^{-1}$ and 15,000 km s$^{-1}$. Table \ref{table1} lists the explosion energy, baryonic and gravitational masses, and 
post-bounce run time. The energies have asymptoted to within a few tens of percent for all models of their final supernova energies
and range from 0.09 to 2.3 Bethes. {The 24-M$_{\odot}$ model is being further scrutinized and is not included here.}
The higher energies are statistically, but not monotonically, associated with more massive progenitors.  The growth of the blast 
energy is depicted in Figure \ref{fig:energy}.  Those models that asymptote early do so at lower energies.  Those models that 
eventually achieve higher explosion energies not only do so later, but experience deeper negative energies for a longer time
before emerging into positive territory.  As described in \S\ref{explosion}, this is what is expected for models with massive (shallow) 
density mantles, if they explode, and these are generally, though not exclusively, for the most massive progenitors ($> 16 {\rm{M}}_{\odot}$ ?). 

The energies shown in Table \ref{table1} and Figure \ref{fig:energy} include the gravitational, thermal, and kinetic energies,
as well as the nuclear reassociation energies, of the ejecta.  They also include the outer mantle binding energies of the as yet unshocked
material.  In this way, all the components of the blast energy are accounted for, except the thermonuclear term.  
The latter could be as much as $\sim$10\% of the total, and will slightly increase our numbers. However, 0.1 M$_{\odot}$ of 
oxygen provides only $\sim$0.1 Bethe, so it is only for the most explosive progenitors with significant 
oxygen and carbon shells for which these mostly gravitation-powered supernovae can have an interesting thermonuclear 
component; this might still be only $\sim$10\%. One would expect that the $^{56}$Ni yields would be higher for the more densely mantled stars, so that 
the thermonuclear energy contribution and $^{56}$Ni yield would be correlated with one another and with progenitor mass\cite{sukhbold:16,2020ApJ...890...51E}. 
Curiously, if the speculations\cite{jones_2016} concerning the possible thermonuclear character of the 
lowest-mass progenitors bear out (though see\cite{2019ApJ...886...22Z}), this correlation might be preserved, though for the other end of the massive-star
mass distribution (``mass function").  Note that the mass function is weighted towards the lower masses. 

\begin{figure*}[ht!]
    \begin{center}
    \includegraphics[width=0.9\linewidth, angle = 0]{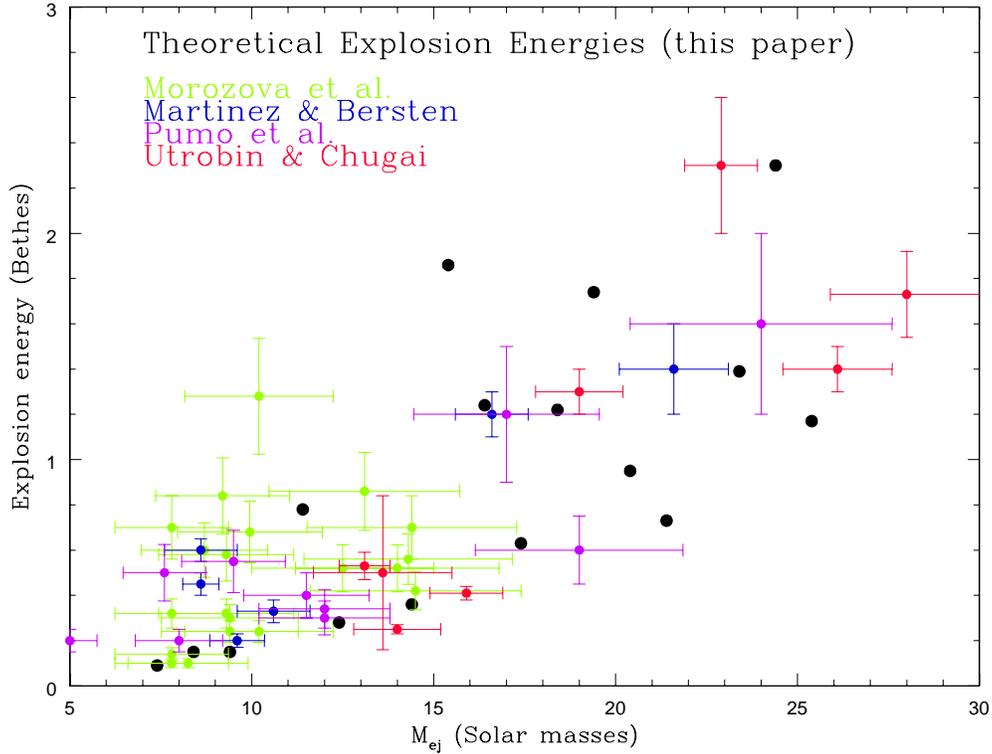}
    \end{center}
    \caption{{\bf Comparison of theoretical and empirical of explosion energy versus ejecta mass.} Plotted are the empirically inferred
explosion energies versus the inferred ejecta masses, with error bars, for a collection of observed Type IIp (plateau) supernovae.  Our theoretical numbers, 
taken from Table \ref{table1}, are superposed as black dots.  It must be recalled that these are 2D models, and that there are
quantitative differences between 2D and 3D simulations. We assume for convenience that the theoretical ejecta masses are the 
progenitor masses, minus the baryon mass of a putative residual neutron star of 1.6 M$_{\odot}$.  This ignores any mass loss prior to explosion,
surely an incorrect assumption by $\sim$1$-$3 M$_{\odot}$. Nevertheless, the rough correspondence between theory and measurement 
is quite encouraging.  Note that the error bars on the measurements are not firm, and do not include any systematic errors
in the light-curve modeling procedures.  In any case, the general average trend from low to high explosion energy from
lower to higher massive-star progenitor mass reflected in the observations is reproduced rather well by the theory, both quantitatively 
and qualitatively.  Note also that at a given mass there is an inferred measured spread in supernova energies.  This may represent
a real variation in explosion energy at a given progenitor mass due in part to the natural chaos in turbulent flow. 
Indeed, it is theoretically expected that Nature would map a given star's properties to distribution functions
in the outcomes and products of its supernova death. The empirical estimates were taken from Morozova et al.\cite{morozova_energy}, 
Martinez \& Bersten\cite{2019A&A...629A.124M}, Pumo et al.\cite{2011ApJ...741...41P,2017MNRAS.464.3013P}, and Utrobin 
\& Chugai\cite{1996A&A...306..219U,2013A&A...555A.145U,2017MNRAS.472.5004U,2019MNRAS.490.2042U}. \label{fig:empirical}}
\end{figure*}

Figure \ref{fig:empirical} superposes the theoretical explosion energies of Table \ref{table1} onto a plot 
of the observationally inferred Type IIp (plateau) supernova energies versus inferred ejecta masses.  For our theory numbers, 
we shift the initial progenitor mass by 1.6 M$_{\odot}$ to account for an average residual neutron star.  In so 
doing, we do not account for the pre-explosion mass loss of the star, which could be significant.  However, the general
trend of the inferred energy with a measure of stellar mass is reproduced by the theoretical (black) dots.  There
is scatter in both the theory and observations, the latter due to systematic uncertainties in the models employed and 
observational limitations, and the former due to numerical and astrophysical uncertainties.  However, natural chaos in 
the dynamics would naturally lead to a spread in energies (\S\ref{intro}, \S\ref{energy}), to a degree as yet unknown, 
even for the same initial stellar structure.   We note that there seems to be a larger observational spread in the 
inferred energies at lower masses.  This could reflect natural chaos in the turbulent neutrino mechanism, measurement 
uncertainties, the effects of unknown rotation, or the possibility that the lowest-mass progenitors explode thermonuclearly
just after the onset of a collapse that does not achieve nuclear densities.  However, it is too soon to draw any 
definitive conclusions on this score. Be that as it may, the observed very roughly monotonically increasing trend of explosion
energy with mass and the ability of the neutrino mechanism to reproduce the observed range of explosion energies are both encouraging. 

Finally, the infalling accretion matter plumes that hit the PNS core generate sound waves that are launched outward.  Much of the energy 
of these sound waves is absorbed behind the shock wave and can modestly contribute to the explosion energy.  Such a component
is automatically include in our bookkeeping. Though difficult to estimate separately, we don't envision that acoustic power 
can contribute more than $\sim$5$-$10\% to the total.

\section{Residual Neutron-Star Masses}
\label{neutron}

\begin{figure*}[ht!]
    \begin{center}
    \includegraphics[width=0.9\linewidth]{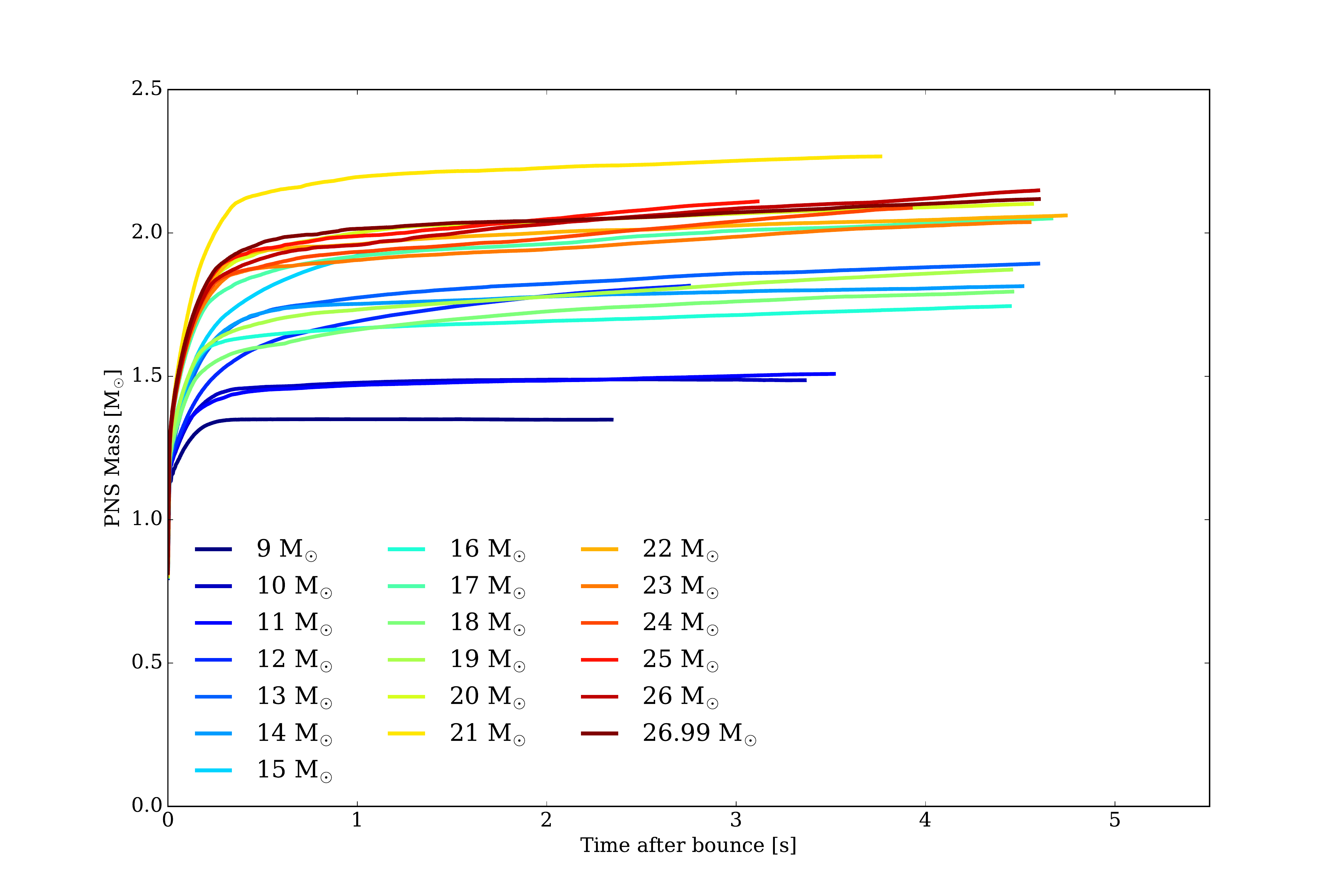}
    \end{center}
    \caption{{\bf Theoretical baryon mass (in M$_{\odot}$) of the residual neutron star versus time after bounce (in seconds)
for the 2D models of this study.} The evolution of the residual neutron star mass is generally rather quick, with 
the final mass determined to within $\sim$5\% generally (though not universally) within $\sim$one second of bounce. The range of 
residual masses ranges from $\sim$1.3 M$_{\odot}$ to 2.2 M$_{\odot}$ for this model set. This is equivalent to a range of 
neutron-star gravitational masses between $\sim$1.2 M$_{\odot}$ and $\sim$2.0 M$_{\odot}$, roughly what is empirically
seen. Generally, the lower-mass progenitors give birth to lower-mass neutron stars, though this is not rigorously monotonic.
Note that the 12-M$_{\odot}$ and 15-M$_{\odot}$ models that don't explode are still gradually increasing their
residual masses by the end of those simulations (see Table \ref{table1}). \label{fig:mass}}
\end{figure*}

Figure \ref{fig:mass} depicts the evolution of the residual baryon mass of the PNS core for the suite of 
2D models investigated here.  Such masses flatten early, since the mass accretion rates drop quickly after 
the explosion commences. The final baryon masses at the last timesteps are given in Table \ref{table1}, as are 
the corresponding gravitational masses. The latter include the gravitational binding energy (negative) 
of the core. These masses range from a low near $\sim$1.2 M$_{\odot}$ to 
a high near $\sim$2.0 M$_{\odot}$, nicely spanning the observed range\cite{2012ApJ...757...55O}.  The neutron star masses we find
are closely, but not perfectly, monotonic with progenitor mass and the shallowness of the Chandrasekhar mantle, 
except for those models that don't explode. Presumably, these models will eventually collapse to black holes, 
but on timescales longer than we have simulated.

\section{Ejecta Compositions}
\label{comp}

The issue of the ejecta elemental composition is fundamental to supernova theory.
The shallowness of the outer mantle density profile and the associated mass of the 
inner ejecta are roughly correlated with the yields of oxygen and intermediate-mass (e.g., Ar, Si, Ca) elements.
As suggested in \S\ref{energy}, such a structure is also likely to explode (if via the 
neutrino mechanism) with higher energies.  Therefore, more of this inner ejecta will be able to 
achieve the higher temperatures that can transform oxygen and silicon into iron-peak species as well.
This includes $^{56}$Ni. Therefore, one expects that in the context of the neutrino mechanism of explosion
$^{56}$Ni yields are roughly increasing functions of progenitor mass, with the exceptions to 
strict monotonicity alluded to previously. Specifically, if a 9-M$_{\odot}$ star explodes 
by the neutrino mechanism it can not have much $^{56}$Ni in its ejecta and if a $\sim$16$-$25 M$_{\odot}$ 
star explodes by the same mechanism the $^{56}$Ni yield should be more significant.

All the inner ejecta from the region interior to the stalled shock wave, before and just after explosion, are very 
neutron-rich ($\rm{Y}_e \sim 0.1-0.2$). As they expand outward, absorption by $\nu_e$ and $\bar{\nu}_e$ 
neutrinos on balance tends to push the ejecta Y$_e$ upward.  If the expansion is fast, then some of the ejecta can freeze out 
slightly neutron-rich below Y$_e = 0.5$.  However, if the expansion is slow, there is plenty of time for 
some of the debris to become proton-rich (Y$_e > 0.5$).  However, generally Y$_e$ = 0.5 seems 
to predominate in the bulk.  Therefore, those models that explode early and fast should provide 
some neutron-rich ejecta, though more of their ejecta could still be proton-rich, while those models that
explode later and more slowly (generally, the more massive progenitors) will be the most proton-rich.
This is what we\cite{2020MNRAS.491.2715B} see, where Y$_e$s from $\sim$0.5 to as high as $\sim$0.58-0.6 
are found.  This might make such supernovae sites for the rp-process and for light p-nuclei 
(e.g., $^{74}$Se, $^{78}$Kr, and $^{84}$Sr)\cite{2006ApJ...644.1028P,frohlich2006,fischer:10}.  However, these numbers should be 
viewed as preliminary, depending as they do on detailed neutrino transport and the complicated trajectory 
histories of the ejecta parcels.  We note that observations of $^{57}$Ni in SN1987A, inferred to be a 
$\sim$18 M$_{\odot}$ progenitor, require that no material with Y$_e$s lower than 0.497 
could have been ejected \cite{1990ApJ...349..222T}.  Also, none of the ejecta seen in modern 
simulations can be the site of all the r-process, though the first peak is not excluded. The 
timescales and Y$_e$s are not at all conducive.

\begin{figure*}[ht!]
    \begin{center}
    \includegraphics[width=1.50\linewidth]{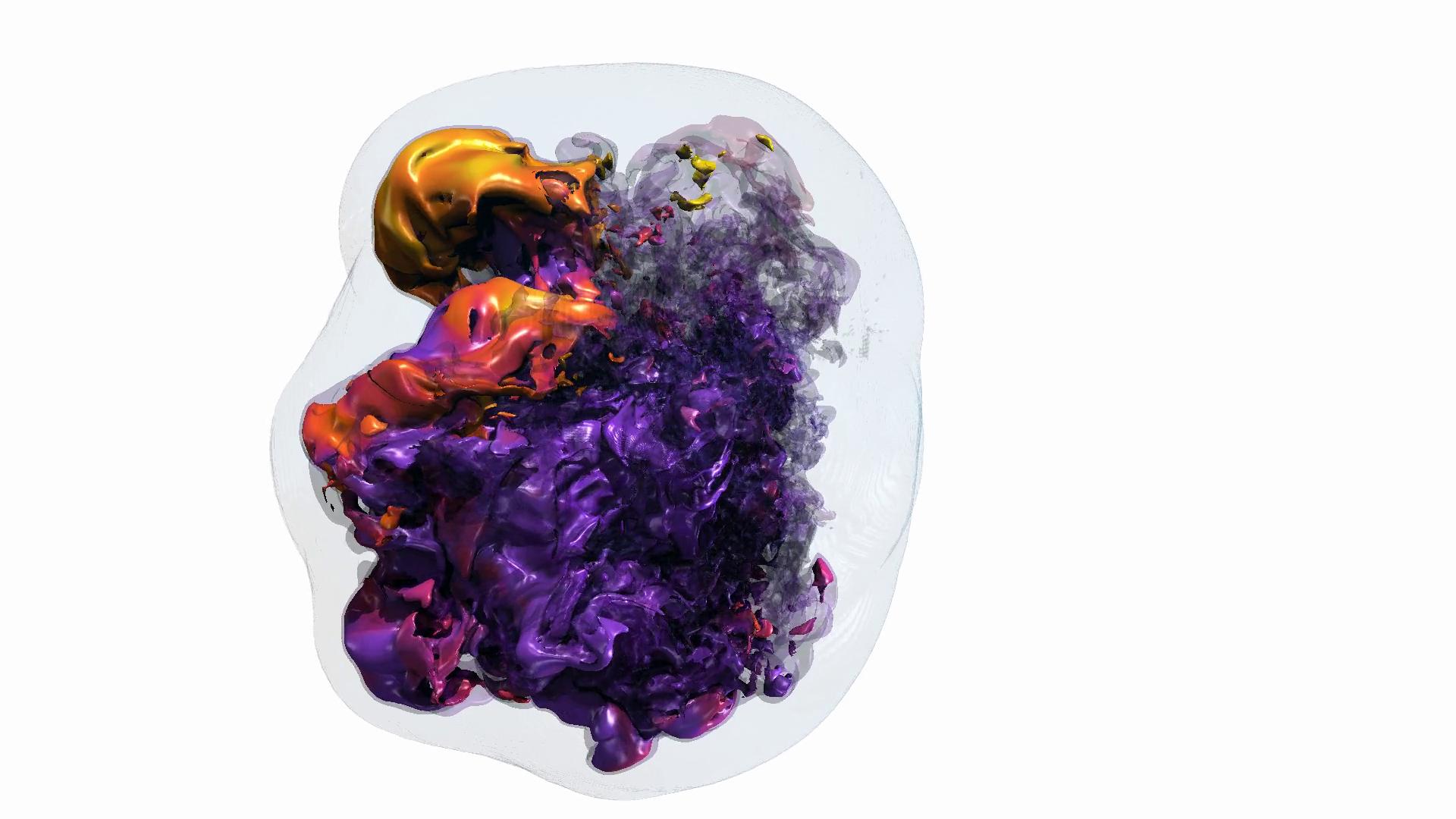}
    \end{center}
    \caption{{\bf 3D Explosion structure of a representative massive star progenitor model.} The associated
simulation was performed with a 678 (radius) $\times$ 256 ($\theta$) $\times$ 512 ($\phi$) grid to render slightly
finer details.  The snapshot is taken $\sim$800 milliseconds after core bounce, about 500 milliseconds into the explosion.
The blue-gray veil is the shock wave.  The colored isosurface is of constant entropy, colored with the electron fraction, Y$_e$.
We note that there is a large region of purple (higher Y$_e$, more proton-rich) matter on one side and a largish region of
orange-yellow (relatively lower Y$_e$, less proton-rich) matter on the other.  This global
Y$_e$ asymmetry is created by persistent angular asymmetries in the electron-neutrino and anti-electron-neutrino
emission from the core during explosion, which by absorption in the ejecta create this asymmetry in the electron
fraction of the ejecta. The latter translates into an asymmetry in the nucleosynthetic element angular distribution.
Results derived from a simulation done by the Princeton supernova group\cite{2019MNRAS.490.4622N}. \label{fig:19solar}}
\end{figure*}

Furthermore, as stated, inner supernova matter explodes quite aspherically, with bubble, botryoidal,
and fractured structures predominating.  However, the spatial distribution of Y$_e$ in the ejecta can have a
roughly dipolar component, with one hemisphere more proton-rich than its counterpart. Figure \ref{fig:19solar}
depicts a snapshot of a simulation of a 19-M$_{\odot}$ model.  The bluish veil is the shock, while the fractured
surface is an isoentropy surface painted by Y$_e$.  As seen, there is an orange-purple dichotomy which reflects
the fact that the ejecta have a dipole in Y$_e$ that persists. Even an initially uniform ejecta Y$_e$
distribution may be unstable to the establishment of such a dipole.  If near and exterior to the $\nu_e$ neutrinosphere
at the ``surface" of the PNS a perturbation in Y$_e$ arises in a given angular patch of the inner ejecta, that perturbation
can grow due the concommitant effect on the absorptive opacity at those angles, which in turn will either suppress
or enhance the $\nu_e$ emissions to push the Y$_e$ evolution of that matter parcel in the same direction.
The progressive diminution of this absorptive Y$_e$ shift effect with distance can freeze the Y$_e$
perturbation. The upshot is then a crudely dipolar distribution in Y$_e$ that tracks a crudely dipolar
angular distribution in the $\nu_e$ and $\bar{\nu}_e$ luminosities and the so-called LESA (``Lepton Emission Sustained Asymmetry")
phenomenon\cite{2014ApJ...792...96T,oconnor_couch2018,2019ApJ...881...36G,2019MNRAS.489.2227V}.
Whether this dipolar asymmetry in Y$_e$ in the ejecta is a generic outcome remains to be seen.

\section{Pulsar Proper Motions - Kicks}
\label{kicks}

The neutron stars born as proto-neutron stars in the supernova cauldron are the source of the radio pulsars known to be 
darting throughout the galaxy with speeds that average $\sim$350 km s$^{-1}$ \cite{2002ApJ...568..289A,faucher_kaspi}
and can range up to $\sim$1500 km s$^{-1}$ \cite{1993Natur.362..133C}. The most natural explanation
for these galactic motions is recoils during the supernova explosion directly related to asymmetric
matter ejection\cite{burrows:96,2004PhRvL..92a1103S,2006A&A...457..963S,wongwathanarat10,2012MNRAS.423.1805N,2013A&A...552A.126W,2019PASJ...71...98N} 
and/or asymmetric neutrino emission. Hence, momentum conservation in the context of at times very aspherical
ejection can easily yield the observed speeds.  Moreover, it is known that neutrino emissions can have a dipolar 
component and that the associated net momentum can be large.  Neutrinos travel at very, very close to the 
speed of light and constitute in sum as much as 0.15 M$_{\odot}$c$^2$ of mass-energy.  Therefore, a mere 1\% 
asymmetry in angle can translate into a kick of $\sim$300 km s$^{-1}$. However, it is not known how the
ejecta and neutrino momentum vectors sum, in particular whether they add or subtract and what the integrated 
magnitude of the latter is. 

Nevertheless, one can speculate about the trends with progenitor star of the magnitude of the kicks experienced\cite{burrows2007_features}.  
We have seen that the lowest-mass massive stars tend to explode a bit more spherically, eject less core mass,
and emit less energy in neutrinos.  The radiated binding energies of the PNS are lower, given the lower accretion rates
and lower PNS mass.  Hence, we expect the kicks to be smaller for the lower mass progenitors.  Conversely,
the more massive progenitors tend to explode a bit more aspherically, ejecting more core mass and emitting 
more mass-energy in neutrinos.  Hence, we posit that they produce neutron stars with the greatest kick speeds.
There is likely to be some noise in these suggestions, but on average these trends with progenitor mass (actually progenitor
structure; see Figure \ref{fig:rho}) are compelling in the context of the neutrino mechanism of CCSN explosions.
Moreover, one would predict that stellar-mass black holes born in the context of core collapse 
would have low kick speeds, since they are generally expected to have much more inertia/mass 
than neutron stars and the momentum in any matter ejecta their birth may entail should be smaller.  
However, the neutrino kicks may be as significant as for neutron-star birth; therefore, the momentum in any such
black hole birth kick might be comparable.

%%%\cite{2020arXiv200608360M}; Mandel and B. M\"uller

\section{Pulsar Spins}
\label{spins}

Stars have angular momentum and spin.  As they evolve, the angular momentum is 
redistributed internally (likely by magnetic torques) and lost in winds.
It is not known what the internal birth spin distribution of massive stars is, but crude theoretical calculations
suggest that angular momentum is gradually transported out of their cores as they
evolve\cite{heger:05}, much of it lost to stellar winds.  In stars for which the internal spin rates can be measured
due to observed rotational splitting of surface pulsational modes, models of the interior
spin evolution leave their interiors rotating ten times too fast\cite{2014ApJ...788...93C}. 
Therefore, the theory of angular momentum redistribution is incomplete. In addition, the cores of massive stars 
shrink and spin up.  So, the spin of a Chandrasekhar core just before collapse is 
a product of the initial angular momentum distribution, wind angular momentum loss, 
redistribution torquing during evolution, and progressive evolutionary compression
of the core.  Furthermore, the spin of the collapsing core can be affected by the stochastic shedding of hydrodynamic waves 
generated during oxygen and silicon core convection just before collapse\cite{2020arXiv200602146M}. 
Without many observational constraints, the final core spins before core collapse are unknown.  

However, radio pulsars have average surface dipole magnetic fields of $\sim$10$^{12}$ gauss and are observed 
on average to be rotating slowly, with average periods of $\sim$500 milliseconds
\cite{1981JApA....2..315V,1987ApJ...319..162N,1989ApJ...345..931E}.  A neutron star needs to be spinning
with a period of $\sim$5 milliseconds to have a rotational kinetic energy of $\sim$10$^{51}$ ergs, so periods
of $\sim$500 ms imply rotational kinetic energies that are four orders of magnitude below supernova 
energies and these are not dynamically important. Nevertheless, the birth spin of a neutron star is an important 
observable and predicting this number should be a goal of theory. 

Even given the spin rate of the Chandrasekhar white dwarf that collapses, 
one can't easily predict the birth spin of the neutron star that eventually emerges. 
Assuming angular momentum conservation, collapsing from this initial configuration 
to a neutron star spins the residue up by approximately a factor of one thousand. So, 
a $\sim$50-second progenitor core translates into a $\sim$50-ms PNS. However, 
the effects of magnetic torques during collapse and after bounce depend on the 
unknown core magnetic fields; whatever the initial B-fields, they are radically affected by subsequent
compression, rotation, turbulence, and various classes of dynamo action. This is particularly true 
after bounce, since the violent turbulence in the PNS is likely to radically change both 
the magnitude and the structure of the magnetic field\cite{1993ApJ...408..194T}. Very large fields
in the magnetar\cite{1992ApJ...392L...9D} range (10$^{15}$ gauss) can spin down the 
nascent PNS on timescales of seconds.    

\begin{figure*}[ht!]
    \begin{center}

    \includegraphics[width=0.95\linewidth]{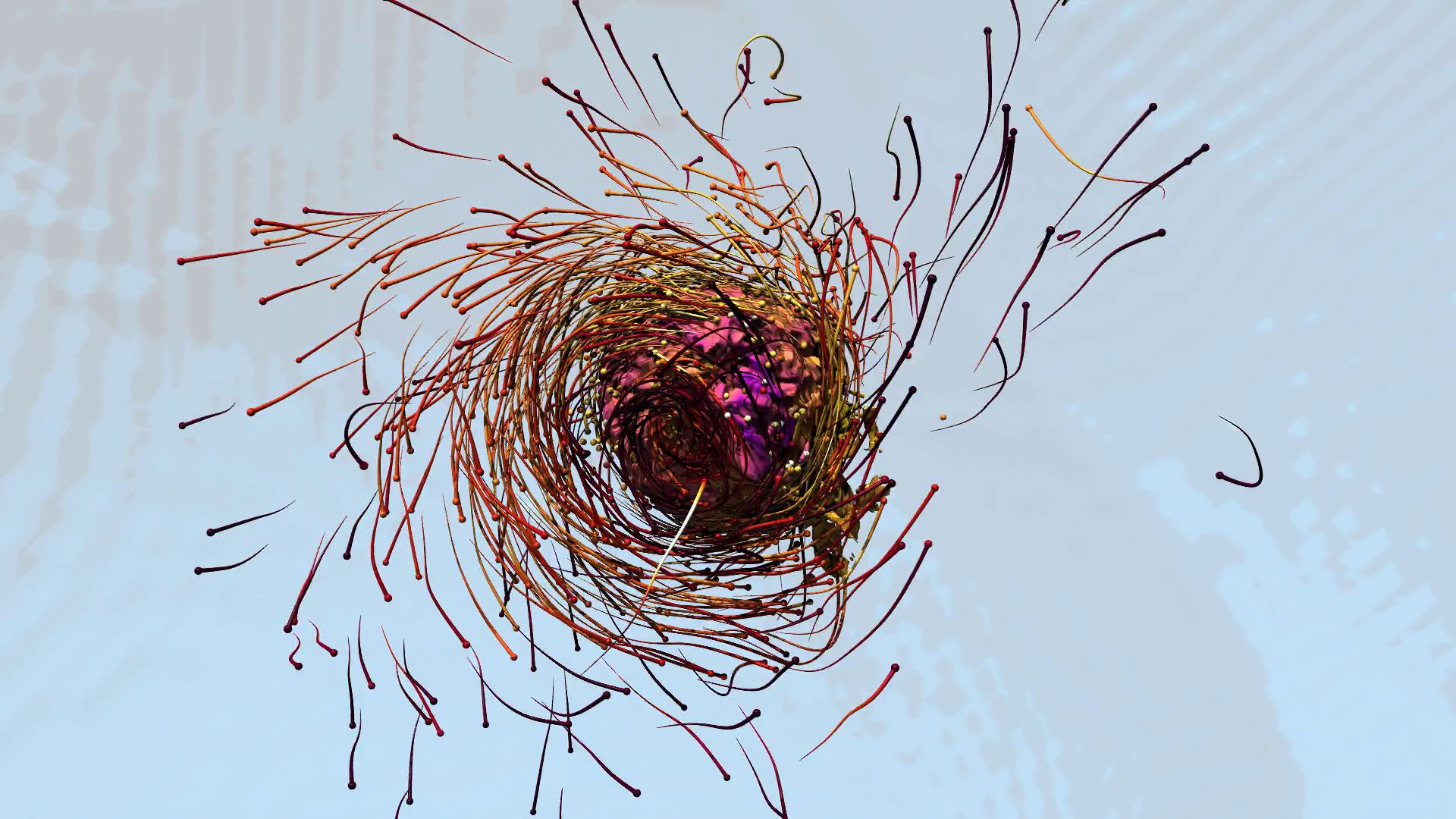}

    \end{center}
    \caption{{\bf Similar to Figure \ref{fig:tracers}, but highlighting the generic swirling
motions just exterior to the PNS a few hundred milliseconds after explosion.}
The physical scale top to bottom is $\sim$200 km. Upon accretion into this inner region, the matter blobs
can stream to one side or another of the core before finally settling onto it.  This stochastic, almost random,
accretion of angular momentum can sum over time to leave a net angular momentum and spin, despite the
fact that the original progenitor model was non-rotating\cite{BlMe07,rantsiou,wongwathanarat13,2020MNRAS.496.2039S}.
\label{fig:rotation}}
\end{figure*}

Furthermore, after bounce the accreting PNS can be spun up by accreting matter plumes 
stochastically, with a jumble of streams with both positive and negative angular momenta that don't necessarily
cancel.  This can lead after the mass cut between the nascent PNS and the ejecta to a 
spinning neutron star, even if the initial star was non-rotating.  
Figure \ref{fig:rotation} depicts emergent rotation at later times. 
The possibility that a previously non-rotating core could be left rotating has been vigorously 
studied\cite{BlMe07,rantsiou,wongwathanarat13}, with a range of spin periods predicted by 
this process alone from a second or two to tens of milliseconds.

Be that as it may, one would like to determine whether the final neutron star spin is predictable.
To date, we don't know.  However, if the total stellar mass and angular momentum loss
are determining factors, one would expect neutron stars born in low-metallicity (low abundance of non-H/He 
elements) massive stars to be faster rotators, all else being equal.  Moreover, if angular transfer from the massive star 
core to its mantle is a continuous process, since the more massive stars evolve more quickly
they are likely to leave cores with more angular momentum, again all else being equal.
So, lower-metallicity, more-massive massive stars would birth neutron stars with faster spins,
but, as implied here, there are still too much uncertainties and too many effectors
to draw reliable conclusions.

\section{Pulsar B-fields}
\label{bfields}

The prejudice of many researchers is that the frozen-in magnetic flux of the unstable Chandrasekhar core
determines the neutron-star fields.  This can't be correct. As stated in \S\ref{spins}, turbulence
behind the shock and in the PNS itself are natural venues for dynamo growth.  At the very least, such violent
convective motions will advect and tangle any initial seed fields post-bounce and the multipolarity structure 
will be radically altered. Even without exponential dynamos, rotation will wind up an initial field and 
the toroidal and poloidal components will evolve significantly.  For large fields near $\sim$10$^{15}$ gauss, 
the field can act back on the rotational profile significantly.  Therefore, it is not at all clear what 
the origin of radio pulsar and magnetar B-fields is, nor what the systematic dependence of these fields might be
on progenitor characteristics;  clearly, this is a rich and important topic for future research and
there have been numerous recent papers attacking aspects of it\cite{2020arXiv200108452M,2020arXiv200306662R,2020MNRAS.492.5764N,2020arXiv200608431F,2020MNRAS.498L.109M}.

A small subset ($\sim$1\% ?) of core-collapse supernova are so-called ``hypernovae" that seem to be missing links with 
long-soft gamma-ray bursts (GRBs)\cite{1993ApJ...405..273W,2010NewAR..54..191N}. Their inferred explosion 
energies are near $\sim$10$^{52}$ ergs ($\sim$10 Bethes) and seem to be too energetic to be powered 
mostly by the neutrino mechanism. The best explanation is that these are powered by MHD jets that tap 
the large spin kinetic energy of fast-spinning (few-millisecond) proto-neutron stars.  Such fast rotators may experience 
strong dynamo action that can achieve magnetar fields.  Hence, the natural consequence of rapid rotation may also be large B-fields,
that together naturally lead to strong jets\cite{burrows2007_mag,mosta2015} that can drive quasi-dipolar outflows.  
As very tentatively suggested in \S\ref{spins}, the progenitors for hypernovae may therefore 
be the more massive massive stars with low-metallicity, and a subset of them may yield 
gamma-ray bursts.  If the latter is the case, a transition within seconds to a black hole as the 
maximum mass of a neutron star is reached is indicated\cite{1993ApJ...405..273W}, since relativistic jets 
are the best explanation for such GRBs and only black holes seem able to produce them.  Therefore, this class of GRBs
would have a non-relativistic jet precursor lasting a few seconds that could be more energetic than the relativistic jet that follows
and would eventually overtake the former as it blasted out of the progenitor star. This non-relativistic/energetic to 
relativistic/less-energetic phasing of jets has yet to be observed in either the hypernova or the GRB context, but is
suggested by emerging theory.

In any case, even if the neutrino mechanism described in this paper predominates among CCSNe, the residual neutron star 
will likely be rotating, however fast, and will have a magnetic field with a dipole contribution.  After the neutrino mechanism
has cleared out the inner cavity around the nascent spinning/magnetic neutron star, this object will be able to transfer power
via weak jets or pulsar action to the inner debris. This may constitute a mere 0.01\% to 1\% of the total supernova energy. 
Eventually, its effects will be manifest in the blast remnant, but perhaps as a sub-dominant, under-energetic phenomenon.  
So, for the canonical case, a standard neutrino-driven explosion could be followed by a weaker magnetically-driven secondary 
effect most of the time\cite{burrows2007_mag,2020ApJ...890...51E}. Signatures of this sort need to be sought, but the supernova 
remnant Cas A, with its sub-dominant jet/counter-jet structure\cite{willingale}, may be such a structure. When the 
initial core is rapidly rotating (we think in a small subset of cases), this magnetically-driven component might 
overtake in energy the neutrino component.  There may also be intermediate cases for which the two mechanisms compete. 
Such a continuum from weak to strong effects of magnetic fields is an intriguing possibility.

\section{Black Hole Formation}
\label{bh}

If and when the PNS mass exceeds the maximum gravitational mass of a neutron star (with suitable small 
thermal and compositional corrections), it will collapse to a black hole and continue to accrete.  
This maximum mass is in the range of 2.1 to 2.4 M$_{\odot}$ gravitational, about 2.4 to 2.7 M$_{\odot}$
baryonic, and depends upon the only modestly-constrained nuclear equation of state.
How much mass is subsequently accreted depends on how much of the progenitor star is ejected.
If none of the star is ejected, and 1) most of the progenitors of such ``stellar-mass" black holes
are the more massive massive stars with high envelope binding energies (\S\ref{explosion}) and 2) these
have experienced significant pre-collapse wind and/or episodic mass loss, then one would expect the 
canonical mass of the product black hole to be $\sim$10-20 M$_{\odot}$\cite{1987PhT....40i..28B}.
This is the helium core mass of those model stars that have very high envelope binding energies exterior 
to the Chandrasekhar core.  However, we\cite{2020MNRAS.491.2715B} had earlier witnessed that stars
with initial masses in the 13$-$15 M$_{\odot}$ range didn't explode.  This result could easily be model-dependent
and is not the final word on what such stars do.  But it is possible that the black-hole outcome is
peppered about the massive-star mass function\cite{1996ApJ...457..834T,2011ApJ...730...70O,sukhbold:16,chan_18,2020arXiv200110492W}.
However, the consensus is that most massive stars with initial masses less than $\sim$20 M$_{\odot}$
will lead to neutron stars and most stars with initial masses greater than $\sim$30 M$_{\odot}$ should lead to black holes.

If most stellar-mass black holes birthed via collapse have masses in the range of $\sim$10 M$_{\odot}$
and neutron stars have a maximum mass a bit above $\sim$2.0 M$_{\odot}$, then there would be a ``mass gap"
between them.  Such a gap is suggested by the data, but is not 
proven\cite{1987PhT....40i..28B,2011ApJ...741..103F}.  It may be that a shock is relaunched,
but has insufficient energy to eject enough of the inner mass and that will then fall back, still launching an 
explosion wave that unbinds the rest of the stellar envelope.  Where this mass cut occurs would determine 
the birth mass of such a black hole.  A ``fallback" black hole is a distinct possibility\cite{chan_18}, 
but may be a small subset of the massive-star mass function.  It is likely that most collapses lead 
to neutron stars, but what the neutron-star/black-hole birth ratio is for the population of massive stars 
is a subject of much current research.

Finally, there are a few points of principle that need to be articulated.  The first is that in the context of 
the collapse of a Chandrasekhar core, it is impossible to collapse directly to a black hole $-$ there
must always be a proto-neutron-star intermediary. This is because the bouncing inner core is out of sonic contact with the 
outer infalling core.  At bounce, the object does not know that it will eventually exceed the maximum mass.
This means that even when a black hole is the final outcome, the PNS core will always have a significant neutrino\cite{shaquann2018} and 
gravitational-wave\cite{2013ApJ...779L..18C} signature.  A signature of the subsequent dynamical 
collapse to a black hole will be the abrupt cessation of both signals\cite{1986ApJ...300..488B}.
Second, given that neutrino energy losses in the range $\sim$0.1-0.4 M$_{\odot}$c$^2$
are ``inevitable," the outer stellar envelope will experience a decrease in the gravitational potential 
it feels.  This will lead to its readjustment on dynamical timescales and the likely the ejection of matter to 
infinity\cite{1980Ap&SS..69..115N,2013ApJ...769..109L}.  Hence, there should always be some sort 
of explosion, even when a black hole forms. Whether it is such a ``potential-shift" explosion, 
one with significant fallback, or one via a disk jet after the black hole and accretion disk form, 
it is difficult to imagine a purely quiescent black hole birth.

\section{Final Thoughts}
\label{final}

As should now be clear, from the vantage of theory, a multitude of effects are of importance 
in determining the viability, character, and strength of a CCSN explosion.  The roles of the initial progenitor
structure; multi-dimensional neutrino radiation transport; general relativity; instabilities, turbulence, and chaos; 
the nuclear interaction and equation of state; neutrino-matter processes and many-body effects; 
resolution and numerical technique; rotation; and B-fields must all be assessed on the road to a resolution 
of this complex problem.  It is this complexity that has paced progress on this multi-physics,
multi-dimensional, and multi-decade puzzle. However, modern theory has grappled with all these issues
and inputs, with the result that state-of-the-art simulations from many groups evince explosions via
the neutrino mechanism with roughly the correct general character and properties.  Not all researchers
agree on the details. nor do they obtain precisely the same results. Nevertheless, to zeroth-order, the 
neutrino mechanism seems to work, and one is, therefore, tempted to declare that the overall problem of the mechanism 
of supernova explosions is solved, the rest being details.  However, these details include the credible mapping 
of progenitor mass and properties to important observables, such as explosion energy, neutron-star mass, 
nucleosynthesis, morphology, pulsar kicks and spins, and B-field magnitudes and multipolarities.  
Chaos will complicate all this, as will remaining uncertainties in microphysics and numerics.  Yet, despite this,
we are confident that core-collapse supernova theory in the year 2020 has reached a milestone, from which it need 
never look back.

%\clearpage

\begin{table*}

\caption{{\bf Explosion Energies and Neutron-Star Masses.} The 9-, 10-, and 11-M$_{\odot}$
progenitors are from the Sukhbold et al. (2016)\cite{sukhbold:16} suite and
were evolved on spherical grids with radial extents of 30,000, 50,000, and
80,000 kilometers (kms), respectively. Progenitors from
12 to 26.99-M$_{\odot}$ were inherited from the Sukhbold et al. (2018)\cite{sukhbold2018} suite.  The
12-, 13-, and 14-M$_{\odot}$ progenitors were evolved on a spherical grid
spanning 80,000 km in radius. All other progenitors were evolved on
spherical grids spanning 100,000 km in radius. All models were evolved in
2D axisymmetry with 1024 radial cells and 128 ($\theta$) angular cells. Thus, there
are some small differences in resolution for the lower-mass progenitors,
where the progenitor grid is truncated at smaller radii so that the temperature
would remain within our equation-of-state table. All models except the 12-
and 15-M$_{\odot}$ progenitors explode. The Run Time quoted is the time after bounce
at nuclear densities.\label{table1}}
\begin{center}
%{\scriptsize
{
%\begin{tabular}{|c|c|c|c|}\hline
\begin{tabular}{ccccc}\hline\hline
\textbf{Model} & \bf{Explosion Energy} & \bf{Run Time}  & \bf{Baryonic
Mass} & \bf{Gravitational Mass}\\
{[{\rm M}$_{\odot}$]} & [B] & [s] & [M$_{\odot}$] & [M$_{\odot}$]\\  \hline
\bf{9}           & 0.09      & 2.34      & 1.35  & 1.23     \\
\bf{10}         & 0.15      & 3.36      & 1.49   & 1.35    \\
\bf{11}         & 0.15
          & 3.52                                                       &
1.51  & 1.37     \\
\bf{12}          & -0.03
           & 2.75                                                       &
1.82   & 1.62   \\
\bf{13}          & 0.78
          & 4.60                                                       &
1.89  & 1.68      \\
\bf{14}          & 0.28
          & 4.51                                                       &
1.81  & 1.62     \\
\bf{15}          & -0.17
           & 1.04                                                       &
1.93  & 1.71       \\
\bf{16}          & 0.36
          & 4.45                                                       &
1.75  & 1.56         \\
\bf{17}          & 1.86
          & 4.66                                                       &
2.05   & 1.81      \\
\bf{18}          & 1.24
          & 4.58                                                       &
1.80      & 1.60        \\
\bf{19}          & 0.63
          & 4.45                                                       &
1.87    & 1.66    \\
\bf{20}          & 1.22
          & 4.56                                                       &
2.10      & 1.85       \\
\bf{21}          & 1.74
          & 3.76                                                       &
2.27    & 1.97    \\
\bf{22}          & 0.95
          & 4.74                                                       &
2.06    & 1.81        \\
\bf{23}          & 0.73
          & 4.55                                                       &
2.04   & 1.80     \\
%\bf{24 }         & 3.3
%           & 3.92                                                       &
%2.09    &  1.84     \\
\bf{25}          & 1.39
          & 3.11                                                       &
2.11   & 1.85   \\
\bf{26 }         & 2.3
           & 4.60                                                       &
2.15   &  1.88      \\
\bf{26.99}       & 1.17                             & 4.60        & 2.12 & 1.86  \\
\hline
\end{tabular}
}
\end{center}
\end{table*}

\clearpage

%\begin{figure*}[ht!]
%    \begin{center}
%%   \includegraphics[width=0.9\linewidth]{figures/mdot.pdf}
%    \includegraphics[width=0.9\linewidth]{fig2.pdf}
%    \end{center}
%    \caption{{\bf Total mass accretion rate onto the core versus time.} Plotted are the mass accretion rates (in M$_{\odot}$ s$^{-1}$)
%versus time after bounce (in seconds) for the whole range of 2D models explored in this study.  The early accretion rates are universally
%high, above a M$_{\odot}$ per second, and then subside over hundreds of milliseconds to much lower values, generally between 0.1 and 0.01
%M$_{\odot}$ s$^{-1}$, and reside near there for many seconds. The later fluctuations reflect the turbulence around the PNS.
%These general histories reflect the initial progenitor density structures almost completely, with the lower-mass progenitors
%experiencing lower accretion rates; they are major determinants of the outcomes of collapse
%and reflect the heterogeneity of the family of supernova progenitors.  We note that though there has been significant progress
%over the years in stellar modeling, these profiles for a given ZAMS (zero-age main-sequence) mass have not yet
%theoretically converged in the literature. Nevertheless, the range of possible density profiles and accretion histories has
%probably been well captured.  \label{fig:mdot}}
%\end{figure*}

\clearpage

\clearpage

\clearpage

\bibliography{sn.update}

\clearpage

\begin{addendum}

\item[Acknowledgements] The authors thank Joe Insley and Silvio Rizzi 
of the Argonne National Laboratory and the Argonne Leadership Computing 
Facility (ALCF) for their considerable support with the 3D graphics. 
The authors acknowledge ongoing collaborations
with Hiroki Nagakura, Davide Radice, Josh Dolence, Aaron Skinner, and Matt Coleman.
They also acknowledge Evan O'Connor regarding the equation of state,
Gabriel Mart\'inez-Pinedo concerning electron capture on heavy nuclei,
Tug Sukhbold and Stan Woosley for providing details concerning the
initial models, and Todd Thompson and Tianshu Wang regarding inelastic scattering.
Funding was provided by the U.S. Department of Energy Office of Science and the Office
of Advanced Scientific Computing Research via the Scientific Discovery
through Advanced Computing (SciDAC4) program and Grant DE-SC0018297
(subaward 00009650) and by the U.S. NSF under Grants AST-1714267 and PHY-1804048 
(the latter via the Max-Planck/Princeton Center (MPPC) for Plasma Physics).
Awards of computer time were provided by the INCITE program using resources of the
ALCF, which is a DOE Office of Science
User Facility supported under Contract DE-AC02-06CH11357, under a Blue Waters 
sustained-petascale computing project, supported by the National Science Foundation (awards OCI-0725070
and ACI-1238993) and the state of Illinois, under a PRAC
allocation from the National Science Foundation (\#OAC-1809073),
and under the award \#TG-AST170045 to the resource Stampede2 in the Extreme 
Science and Engineering Discovery Environment (XSEDE, ACI-1548562).  Finally,
the authors employed computational resources provided by the TIGRESS high
performance computer center at Princeton University, which is jointly
supported by the Princeton Institute for Computational Science and
Engineering (PICSciE) and the Princeton University Office of Information
Technology, and acknowledge their continuing allocation at the National
Energy Research Scientific Computing Center (NERSC),
supported by the Office of Science of the US Department of Energy
(DOE) under contract DE-AC03-76SF00098.

\item[Author contributions] A.B. organized the paper and wrote most of it.  D.V. conducted 
the 2D calculations. Otherwise, the authors contributed equally to the document.

\item[Additional information]
{\bf~\\Supplementary information} is not available for this paper.
{\bf~\\Reprints and permissions information} is available at \url{www.nature.com/reprints}.
{\bf~\\Correspondence and requests for materials} should be address to A.B.

%% Adjust the following as necessary
\item[Competing interests] The authors declare no competing financial interests.

\end{addendum}

\end{document}